\documentclass[11pt]{article}

\usepackage[]{graphics,graphicx}

\usepackage{pstricks}

\usepackage{blindtext}

\usepackage{setspace} 

\RequirePackage{lineno}

\begin{document} 

 \modulolinenumbers[1]


\begin{center}{\LARGE\bf The Mechanics of Landslide Mobility with Erosion}
\\[10mm]
{Shiva P. Pudasaini$^{\mbox{\,a,b}}$, Michael Krautblatter$^{\mbox{\,a}}$
\\[3mm]
{$^{\mbox{a\,}}$Technical University of Munich, Chair of Landslide Research}\\
{Arcisstrasse 21, D-80333, Munich, Germany}\\[3mm]
{$^{\mbox{b\,}}$University of Bonn,
Institute of Geosciences, Geophysics Section\\
Meckenheimer Allee 176, D-53115, Bonn, Germany}\\[1mm]
{E-mail: shiva.pudasaini@tum.de}\\[7mm]
}
\end{center}
\noindent
{\bf Abstract:} 
Erosion, as a key control of landslide dynamics, significantly increases the destructive power by rapidly amplifying its volume, mobility and impact energy. Mobility is directly linked to the threat posed by an erosive landslide. No clear-cut mechanical condition has been presented so far for when, how and how much energy the erosive landslide gains or loses, resulting in enhanced or reduced mobility. We pioneer a mechanical model for the energy budget of an erosive landslide that controls enhanced or reduced mobility. A fundamentally new understanding is that the increased inertia due to the increased mass is related to an entrainment velocity emerging from the inertial frame of reference. With this the true inertia of an erosive landslide can be ascertained, and its control on the landslide mobility. This makes a breakthrough in correctly determining the mobility of the erosive landslide. Erosion velocity plays an outstanding role in regulating the energy budget and decides whether the landslide mobility will be enhanced, reduced or remains unaltered. This depends exclusively on whether the energy generator is positive, negative or zero. This provides the first-ever explicit mechanical quantification of the state of erosional energy and a precise description of mobility. This addresses the long-standing scientific question of why many erosive landslides generate higher mobility, while others reduce mobility. By introducing three key mechanical concepts: erosion-velocity, entrainment-velocity and energy-velocity, we demonstrate that the erosion and entrainment are essentially different processes. Landslides gain energy and enhance mobility if the erosion velocity is greater than the entrainment velocity. The energy velocity delineates the three excess energy regimes: positive, negative and zero. We introduce two dimensionless numbers, the mobility scaling and erosion number, delivering an explicit measure of mobility. We establish a mechanism of landslide-propulsion providing the erosion-thrust to the landslide. Analytically obtained velocity indicates the fact that erosion can have the major control on the landslide dynamics. To prepare enhanced modelling of entrainment-related mobilization, we also present a full set of dynamical equations in conservative form in which the momentum balance correctly includes the erosion-induced net momentum production.

\section{Introduction}

Erosion, entrainment and deposition are dominant and complex mechanical
processes in geophysical mass flows including landslides, avalanches and
debris flows. Such events can significantly to dramatically increase their volume and
destructive potential, and become exceptionally mobile by entraining sediment from the bed as they rush down mountain slopes (Huggel et al., 2005; Hungr et al., 2005; Santi et al., 2008; de Haas and van Woerkom, 2016; Mergili et al., 2018, 2020b). 
Landslide mobility is associated with erosion-induced excessive volume and material properties, and is characterized by enormous impact energy,
 exceptional travel distance and inundation (coverage) area. Mobility is among the most important features of the landslide as it directly measures the threat posed by the landslide. Mobility is governed by the state of energy of the landslide
 and entrainment can increase the landslide volume by several orders of magnitude
(Evans et al., 2009; Le and Pitman, 2009; Theule et al., 2015; de Haas and van Woerkom, 2016; Liu et al., 2019). The breach of the
moraine dam of Lake Palcacocha by the 1941 glacial lake outburst flood
event (Cordillera Blanca, Peru) lowered the valley bottom by as much as
50 m in some parts (Somos-Valenzuela et al., 2016). As erosion-induced excessive volume is the prime control on the flow dynamics including velocity, travel distance, 
flow depth and impact area, as well as on the number of fatalities (Huggel et al., 2005; Evans et al., 2009; Le and Pitman, 2009; Dowling and Santi, 2014; de Haas et al., 2020), 
erosion, entrainment and associated flow bulking in landslide prone areas and debris-flow torrents are a major concern for civil and environmental engineers and landuse planners. 
Estimations of
flow volume, velocity and the travel distance are key for assessment of mass flow hazard, design of
protective structures and mitigation measures (Rickenmann, 1999; Dowling
and Santi, 2014; de Haas et al., 2020).
 Different field and laboratory studies on bed sediment entrainment (Egashira et
al., 2001; Rickenmann et al., 2003; Hungr et al., 2005; Berger et al., 2011; Iverson et al., 2011; Reid et al.,
2011; McCoy et al., 2012) have suggested that the spatially varying erosion rates and entrainment
processes are dependent on the geomorphological, lithological and mechanical 
conditions. These processes are of theoretical and practical 
interests both for scientists and engineers (Dietrich and Krautblatter, 2019). A proper understanding of
landslide erosion, entrainment and resulting increase in mass (or, volume) is a basic
requirement for an appropriate modelling of landslide motion and its impact
because the associated risk is directly related to the landslide
mass and its velocity. However, as the mechanical controls of erosion and
entrainment have not been well understood yet, evolving volume, run-out and
impact energy of landslides and debris flows are often largely
underestimated (Dietrich and Krautblatter, 2019). 
\\[3mm]
Physical experiments (Iverson et al., 2011; de Haas and van Woerkom,
2016; Lu et al., 2016; Lanzoni et al., 2017, Li et al., 2017) and
theoretical modelling (Fraccarollo and Capart, 2002; Le and Pitman, 2009; Iverson and Ouyang, 2015; Pudasaini and Fischer, 2020) demonstrate the importance of erosion phenomena in landslides and debris
flows.
 In the recent years, there has been rapid increase in the studies of erosion and
entrainment in both laboratory (Fraccarollo and Capart, 2002; Iverson et al.,
2011; de Haas and van Woerkom, 2016) and field scales (Cascini et al., 2014, 2016; Cuomo et al., 2014,
2016; de Haas et al., 2020). Hereby, continuum mechanical (Jenkins and Berzi, 2016; Pudasaini and Fischer, 2020) and kinetic
theory (Berzi and Fraccarollo, 2015) approaches have been applied to
investigate erosion phenomena. Empirical (Takahashi and Kuang, 1986; Rickenmann et al., 2003; McDougall and Hungr, 2005; Chen
et al., 2006;  Le and Pitman, 2009) and
mechanical (Fraccarollo and Capart, 2002; Iverson, 2012)
erosion models have been developed. Most erosion models consider
effectively single-phase flows (Fraccarollo and Capart, 2002; Naaim et al., 2003; McDougall and Hungr,
2005; Tai and Kuo, 2008; Le and Pitman, 2009; Armanini et al., 2009;
 Iverson, 2012). Erosion may depend on the flow depth, flow velocity, solid concentration,
density ratio, bed slope or, the effective stresses at the interface, and
initial and boundary conditions (Brufau et al., 2000; Fagents and Baloga, 2006; Sovilla et al.,
2006; Crosta et al., 2009; Berger et al., 2011).
 Recently, there have been a rapid increase in the number of numerical
models incorporating erosion (McDougall and Hungr, 2005; Medina et al.,
2008; Le and Pitman, 2009; Christen et al., 2010; Frank et al., 2015; Iverson and Ouyang, 2015). However, the erosion
rates presented and utilized in these works are either not based on the
physical principles or, are physically inconsistent (de Haas et al., 2020;
Pudasaini and Fischer, 2020).
\\[3mm]
Although erosion and deposition play important role in mass transport and
shaping the landscape (Huggel et al., 2005; Evans et al., 2009; Dietrich and Krautblatter, 2019; de Haas et al., 2020; Mergili et al., 2020a,b) our understanding of these
processes is not sufficient to apply them beyond empirical experience. Despite the importance of entrainment to hazard
assessment and landscape evolution (Cascini et al., 2014, 2016; Cuomo et
al., 2014, 2016), a clear understanding of the basic process still remains
elusive owing to a lack of high-resolution field-scale data, and also limited flow parameters in
laboratory experiments (de Haas and van
Woerkom, 2016). 
However, due to the complex terrain, infrequent occurrence, and
high time and cost demands of field measurements, the available field data
(Berger et al., 2011; Sch\"urch et al., 2011; McCoy et al., 2012; Theule
et al., 2015; Dietrich and Krautblatter, 2019) are insufficient (de Haas et
al., 2020).
 This is because a proper understanding and interpretation of the data obtained from the
field measurements is often challenging because of the very limited
nature of the material properties of the flow and the boundary conditions.
 Measurements are locally or discretely based
on points in time and space (Berger et al., 2011; Sch\"urch et al., 2011;
McCoy et al., 2012; Theule et al., 2015; Dietrich and Krautblatter, 2019). 
Physics-based models and numerical simulations may
overcome these limitations and facilitate a more complete
understanding by investigating much wider aspects of the flow parameters,
erosion, mobility and deposition.
Similarly, exact, analytical solutions (Faug et al., 2010; Pudasaini, 2011; Pudasaini et al., 2018; Pilvar et al., 2019) can provide
important insights into the complex flow behaviors and their consequences.
\\[3mm]
By extending the general debris flow model (Pudasaini, 2012), 
Pudasaini and Fischer (2020) proposed a novel and process-based two-phase
erosion-deposition model, which, to a large extent, is capable of adequately describing these complex
phenomena commonly observed in landslides, avalanches, debris flows and
bedload transports. These mechanical erosion-rate models proved that the
effectively reduced friction (force) in erosion is equivalent to the momentum
production. This solves the long standing dilemma of mass
mobility, and show that erosion can enhance the mass flow mobility.
The importance of the mechanical erosion model for two-phase mass flows
consisting of viscous fluid and solid particles presented in Pudasaini and
Fischer (2020) has been increasingly realized in the recent simulations of
mixture mass flows including the real events (Li et al., 2019; Qiao et
al., 2019; Shen et al., 2019; Liu and He, 2020; Mergili et al., 2020a,b; Nikooei and
Manzari, 2020; Liu et al., 2021). These modelling approaches have clearly indicated how the mechanical erosion model (Pudasaini and Fischer, 2020) could appropriately simulate the actual flow dynamics, surge development,
run-out or mobility, and deposition morphology based on the mechanical erosion rates
 and the erosion-induced momentum productions.
\\[3mm]
Bed entrainment plays a major role in determining the whole propagation pattern. However, there
can be different aspects of erosion in nature (Pudasaini and Fischer,
2020). 
Considering two Italian 1998 Sarno-Quindici events, Cascini et al. (2014)
mentioned that increasing entrainment rate inside the channel may diminish
the final run-out of channelized landslides of the flow type. With
reference to the 2005 Nocera Inferiore debris avalanche and the 1999
Cervinara debris avalanche, Cuomo et al. (2014) indicated that bed
entrainment can be a dissipative mechanism to reduce mobility of
unchannelized flow-like landslides. Furthermore, Cuomo et al. (2016)
investigated the role of bed entrainment for the field-observed scenario
of several channelized and unchannelized flows interacting during the
propagation, and concluded that bed entrainment in the central-lower part
of the propagation path could have reduced the run-out. Pudasaini and
Fischer (2020) explained these findings as follows: Because of the
entrained material from the ground to the moving mass, the
reduction in kinetic energy (or, the momentum) of the system might be
greater than the increase in its potential energy. This can probably be
the situation, e.g., for flows in moderate to low slope angles.
Nevertheless, the erosion-related mobility can be site and material
specific. However, those complex behaviors also depend on the type of erosional
mechanism and the mechanisms of momentum exchange, the involved flow rheologies, and how the
net mass and momentum productions are considered in the dynamical model
equations.
The Pudasaini and Fischer (2020) model built a foundation by mechanically including the momentum production. Nevertheless, their model appear incomplete as they did not deal with the inertia of the entrained mass and could not present a clear mechanical condition for when and how the mobility of an erosive landslide will be enhanced or reduced and how to quantify it. 
\\[3mm]
Thus, whether erosion will result in the enhanced mass flow mobility and the quantification of its influence remains an unsolved problem. Extending the Pudasaini and Fischer (2020) model, we have addressed this important issue by explicitly deriving mechanical conditions for the mobility of erosive landslides. By introducing a simple landslide mobility equation, we mechanically explain how and when erosive landslides  enhance or reduce their mobility. 
This has been made possible by physically correctly considering the inertia of the erosive landslide. 
The model offers the first-ever opportunity to distinctly quantify the mobility of an erosive landslide. We have also presented some analytical results with plausible parameters and revealed several major novel dynamical aspects associated with erosion-induced landslide mobility. 
We revealed that the 
 erosion velocity plays an outstanding role in appropriately determining the energy budget of an erosive landslide, providing a precise description of mobility in terms of energy velocity and energy generator. Importantly, a novel mechanism of landslide-propulsion has been identified that emerges from the net momentum production, providing the erosion-thrust to the landslide. 
We constructed two dimensionless numbers, the mobility scaling and the erosion number, the first of this kind in mass flow modelling, delivering the explicit measure of mobility. The mobility scaling (precisely) quantifies the contribution of erosion in landslide mobility.
\\[3mm]
We analytically obtained the landslide velocity by solving the landslide mobility equation that quantifies the effect of erosion in landslide mobility. The explicit form of velocity is technically important in solving engineering problems as it provides the practitioners with key information in quickly estimating the impact force for erosive landslide. Obtained velocities are representative of natural landslides with erosion and indicate the fact that erosion can have the major control on the landslide dynamics. We have also derived a full set of dynamical equations in conservative form, in which the momentum balance correctly includes the erosion-induced change in inertia and the momentum production. This constitutes a foundation for legitimate and physically plausible simulation of landslide motion with erosion. 

\section{Basic Balance Equations for Erosive Landslide}

\subsection{Mass and momentum balance equations}

For simplicity, we consider a geometrically two-dimensional flow. The superscripts $^m$ and $^b$ represent the parameters or the quantities in the landslide (mixture), and the erodible basal substrate, respectively. Let $t$ be the time, $(x,
z)$ be the coordinates and $\left ( g^x, g^z\right )$ the gravity
accelerations along and perpendicular to the slope, respectively. Let, $h$ and $u$ be the
flow depth and the mean flow velocity along the slope. Similarly,
$\gamma^m, \alpha^m, \mu^m$ be the density ratio between the fluid and the
particles $\left ( \gamma^m = \rho^m_f/\rho^m_s\right )$, volume fraction of the solid
particles (coarse and fine solid particles), and the basal friction coefficient $\left ( \mu^m =
\tan\delta^m\right )$ related to the basal friction angle $\delta^m$, in
the mixture. Furthermore, $E$ is the basal erosion rate, $u^b$ is
the velocity of the eroded material from the basal substrate, $K$ is the earth pressure coefficient as a function of
 the internal $(\phi^m)$ and the basal $(\delta^m)$ friction angles, and $C_{DV}$ is the viscous drag coefficient.
\\[3mm]
Consider the multi-phase mass flow model (Pudasaini and Mergili, 2019), and
include the viscous drag and erosion (Pudasaini and Fischer,
2020). 
For simplicity, we assume that the relative velocity between the coarse and fine solid
particles $(u_s, u_{fs})$ and the fluid phase $(u_f)$ in the debris material is negligible, that
is, $u_s \approx u_{fs} \approx u_f = u$, and so is the viscous deformation of the fluid.
This means, to facilitate for the derivation of a simple model, we are considering an effectively single-phase mixture flow.
Then, by summing up the mass and momentum balance equations, we obtain a single mass and
momentum balance equation describing the motion of an erosive landslide as:
\begin{equation}
\frac{\partial h}{\partial t} +  \frac{\partial }{\partial x}\left ( hu\right ) = E,
\label{Eqn_1}
\end{equation}
\begin{eqnarray}
&&\frac{\partial}{\partial t}\left( hu \right)+ \frac{\partial}{\partial x} \left [ hu^2 + \left ( 1 - \gamma^m\right)\alpha^m g^z K\frac{h^2}{2}\right] \nonumber\\
&&= h \left[g^x - \left ( 1-\gamma^m \right ) \alpha^m g^z\mu^m
-g^z\left \{ 1- \left( 1-\gamma^m\right )\alpha^m \right \} \frac{\partial h}{\partial x} - C_{DV} u^2\right] + u^b E,
\label{Eqn_2}
\end{eqnarray}
where $E$ and $u^b E$ are the mass and momentum productions, respectively, and $-\left ( 1-\alpha^m\right)g^z\partial h/\partial x$ emerges from the hydraulic pressure gradient associated with the possible interstitial fluid in the landslide. Moreover, the term containing $K$ on the left hand side, and the right hand side in the momentum equation (\ref{Eqn_2}) represent all the involved forces. 
The first term in the square bracket on the left hand side of (\ref{Eqn_2}) describes the advection, while the second term describes the extent
of the local deformation that stems from the hydraulic pressure gradient of the free-surface of the flow. The first, second, third, fourth and the fifth
terms on the right hand side of (\ref{Eqn_2}) are forces due to the gravity acceleration; effective Coulomb friction
which includes lubrication due to the buoyancy $\left (1-\gamma^m\right)$, liquefaction due to the
solid volume fraction $\left ( \alpha^m\right )$; the term associated with buoyancy and the fluid related hydraulic pressure gradient; viscous drag and the erosion
induced momentum production, respectively. By setting $\gamma^m = 0$ and $\alpha^m = 1$, we obtain a pure dry granular flow or an avalanche motion. For this choice, the third term on the right hand side vanishes. However, we keep $\gamma^m$ and $\alpha^m$ also to include different aspects of possible fluid effects in the landslide. 
\\[3mm]
The momentum balance equation (\ref{Eqn_2}) can be re-written as: 
\begin{eqnarray}
&&h\left[\frac{\partial u}{\partial t} + u \frac{\partial u}{\partial x}\right]
+ u\left[\frac{\partial h}{\partial t} + \frac{\partial}{\partial x}\left ( hu\right )\right]\nonumber\\
 &&= h\left[g^x - (1-\gamma^m)\alpha^m g^z\mu^m 
 - g^z
 \left \{ \left( \left( 1-\gamma^m\right)K + \gamma^m\right)\alpha^m+\left ( 1-\alpha^m \right )\right \}
 \frac{\partial h}{\partial x}
 -C_{DV}u^2 \right]+ u^b E.  
 \label{Eqn_3}
 \end{eqnarray}
Note that for $K = 1$ (which mostly prevails for extensional flows, Pudasaini and Hutter, 2007), the third term on the right hand side associated with $\partial h/\partial x$ simplifies drastically, because $\left \{ \left( \left( 1-\gamma^m\right)K + \gamma^m\right)\alpha^m+\left ( 1-\alpha^m \right )\right \}$ becomes unity.
 This also indicates that by assuming the isotropic mixture ($K = 1$) one loses some important information about the solid content and the buoyancy effect.

\subsection{Eliminating existing erroneous perception on erosive landslide,\\ and making a breakthrough}

As entrainment introduces new mass into the system the inertia is increased. One might simply think that the expression $\displaystyle{u\left[\frac{\partial h}{\partial t} + \frac{\partial}{\partial x}\left ( hu\right )\right]}$ on the left hand side in the momentum equation (\ref{Eqn_3}) can be replaced by $uE$, where the flow velocity $u$ is multiplied by the erosion rate (the time rate of change of mass resulting in the mass production), $E$ from the mass balance (\ref{Eqn_1}). However, here, one must be very cautious. The velocity associated with this increased mass must be handled carefully. In reality, the erosion-induced produced mass (with the rate $E$) is not transported by the flow velocity $u$ itself but, it is transported by a fundamentally different velocity that can be substantially lower than $u$. As we will see later, this newly revealed fact becomes a game-changer and makes a breakthrough in correctly determining the state of energy (or, momentum) and thus the mobility associated with an erosive landslide. Below, we derive a physically correct momentum balance equation for an erosional landslide and prove that the direct substitution $\displaystyle{ u \left[\frac{\partial h}{\partial t} + \frac{\partial}{\partial x}\left ( hu\right )\right]=uE}$ in the inertial part of the momentum balance equation (\ref{Eqn_3}) is physically wrong. This  appears from an erroneous understanding of the erosional landslide, but prevails in many existing models (see, e.g., Perla et al., 1980; McDougall and Hungr, 2005; Iverson, 2012).

\section{Correct Derivation of the Relevant Momentum Balance Equation}

\subsection{Basic erosional landslide equation}

We derive a basic erosional landslide equation in the most elegant way. The situation of an erosive landslide is as follows. Let $m+\left ( -\Delta m\right )$ be the mass of the landslide that moves with velocity $u$ at time $t_1 = t$. Then, at time $t_2 = t+\Delta t$, after entraining the mass $\Delta m$, the actual landslide of mass  $m$ moves with velocity $u + \Delta u$, and the mass $-\Delta m$ moves with the erosion velocity $u^b$.
\\[5mm]
So, the momentum $P_1$ of the landslide at time $t_1$, and the momentum $P_2$ of the landslide and the eroded mass at time $t_2$, respectively, are:
\begin{equation}
P_1 = \left [ m + \left ( -\Delta m\right )\right ] u, 
\label{Eqn_4}
\end{equation}
and 
\begin{equation}
P_2 =\left [ m + \left ( -\Delta m\right ) + \Delta m\right ]\left ( u + \Delta u\right ) + \left ( -\Delta m\right )u^b
= m\left( u + \Delta u\right) + \left ( -\Delta m\right )u^b. 
\label{Eqn_5}
\end{equation}
The form of $P_2$, particularly the appearance of the negative sign in
$(-\Delta m) u^b$ in (\ref{Eqn_5}), may seem to be abnormal. In fact, the negative
sign is not due to the negative change of the mass but rather due to the
relative erosion velocity as compared to the actual velocity of
the landslide. This can be proven rigorously in many different but
equivalent ways. We present here one of the best ways: In time $t_2$ the
mass $(m - \Delta m)$ moves with velocity $(u + \Delta u)$, and from the
frame of reference of the landslide, the entrained mass $\Delta m$ moves
with the velocity $(u + \Delta u) – u^b$. These two components in $P_2$
can be re-arranged as follows:
\begin{eqnarray}
P_2
&=& (m - \Delta m)( u + \Delta u) +  \Delta m \left[(u + \Delta u) – u^b\right]
= [(m - \Delta m)( u + \Delta u) + \Delta m (u + \Delta u)] + \Delta m(-u^b)
\nonumber\\
&=& m (u + \Delta u) + \Delta m \left(-u^b\right)
=  m (u + \Delta u) + \left(– \Delta m\right)u^b.
\end{eqnarray}
So, looking on the nicely selected form of the momentum $P_1 = [m
+\left(- \Delta m\right)] u$ at $t_1$, we can legitimately separate the mass into
two elements $m$ and $\left(- \Delta m\right)$ and assign them
respectively to the velocities $( u + \Delta u) $ and $ u^b $ for the text
time $t_2$ to obtain $P_2$ in (\ref{Eqn_5}). This shows the graceful
consideration in deriving (\ref{Eqn_5}).
\\[3mm]
Conservation of linear momentum states the following relation incorporating all the forces $F$ including the forces applied to the landslide and the entrained mass:
\begin{equation}
    F = \lim_{\Delta t \to 0} \frac{P_2 - P_1}{\Delta t}.  
    \label{Eqn_6}
\end{equation}
Since $P_2 - P_1 = m \Delta u + \left ( u - u^b\right )\Delta m$, we have now the formally and correctly derived momentum equation for an erosional landslide:
 \begin{equation}
F = m\frac{du}{dt} + u^{ev} \frac{dm}{dt},   
\label{Eqn_7}
\end{equation}
where, $u^{ev} = u - u^b $ is the velocity of the eroded mass in the frame of reference of the landslide, which is the relative velocity of the eroded mass with respect to the velocity of the landslide. Moreover, $dm/dt$ is positive. We call (\ref{Eqn_7}) the (basic) erosional landslide equation. 
Equation (\ref{Eqn_7}) can be obtained in many different ways.
One may derive a similar equation for the depositional landslide. 
\\[3mm]
The fundamental understanding here, as revealed by (\ref{Eqn_7}) is, that the increased inertia due to the increase in the mass of the landslide is not related to the velocity of the landslide $u$, but it is associated with the entrainment velocity in the inertial frame of reference of the landslide, $ u^{ev}$. Depending on the erosion situation, that we will discuss later, $ u^{ev}$ can be substantially less than the landslide velocity $u$. Thus, the true increased inertia $ u^{ev}\,{dm}/{dt}$ can be much less than incorrectly proposed previously, $ u\,{dm}/{dt}$. Hence, the classical, or the direct representation of $\displaystyle{ F = \frac{d}{dt}\left ( m u\right)}$ as 
$\displaystyle{ F = m\frac{du}{dt} + u\frac{dm}{dt}}$ is fundamentally wrong for erosional situation. This led many to the erroneous conclusion: 
that either erosion results only in reduced mass flow mobility, because the  landslide consumes more energy resulting in the reduced mobility of the erosive landslide, or that erosion does not change the mass flow mobility as the energy loss in entrainment is balanced by the produced momentum. Later, we prove that, in general, both conclusions are mechanically incorrect. 
\\[3mm]
In special situation, when the eroded mass enters the landslide with almost the velocity of the landslide itself, then $u^{ev} \approx 0$, and there is (almost) no increase (change) of inertia. This can happen if the basal substrate is very weak. Examples include a fully saturated or, liquefied bed material (Iverson, 2012) such that with almost no consumption of energy, the basal substrate can be eroded. However, in a particular situation if the substrate is so strong mechanically that the erosion hardly takes place, and even if it takes place, the erosion velocity $\left ( u^b\right)$ can be as low as zero, only then, the classical approach might seem to be applicable. Which, effectively means, that the classical approach works only for non-erosional situation, but not for landslide with erosion. So, those models, which are based on the unphysical formulation of the momentum balance, are not appropriate in simulating landslide motion with erosion.
\\[3mm]
This implies, that the correct consideration of the inertial frame of reference is crucial for the precise derivation of the dynamical landslide model with erosion. However, it is evident, that the law (\ref{Eqn_7}) cannot be obtained directly by rearranging the inertial terms in the Newton’s second law of motion, but rather must be derived carefully by correctly considering the conservation of momentum for an erosional landslide as done above. Those erosion models that are based on the direct use of the Newton’s second law of motion with regard to the inertial part of the momentum balance equation cannot represent the true mechanism of erosion and the subsequent dynamics. 

\subsection{The landslide-rocket-equation}

In the form, (\ref{Eqn_7}) is similar to the famous Tsiolkovsky Rocket-Equation (Sokolsky, 1968). However, there are fundamental differences. First, the way we derive the model is different. We elegantly considered the mass in time $t$ that has legitimately been split into the landslide mass minus the mass that will later be added into the landslide as the eroded mass at time $t+\Delta t$. This was vital. Second, the mass of the rocket is decreasing (since it consumes fuel), so $dm$ is negative. But, for erosional landslide $dm$ is positive as the mass of landslide is increasing. Third, although $ u^{ev}$ is positive for both the erosional landslide and the rocket, they have quite different perspectives on their magnitudes. For the rocket, it is the velocity of the rocket plus the velocity of the exhaust (because of the negative direction of the exhaust). So, it is more than the (instantaneous) velocity of the rocket. But, for the erosional landslide, it is the velocity of the landslide minus the velocity of the eroded mass that is entrained by the landslide. Thus, depending on the magnitude of the erosion velocity, the entrainment velocity $ u^{ev}$ can be substantially less than the landslide velocity, as the velocity of the eroded particle, that is entrained by the landslide, is a positive quantity that, depending on the situation (the flow and the bed morphology), can be as high as the velocity of the landslide itself.    

\section{The Landslide Mobility Equation: A Novel Model Formulation}

Since, in general, the erosion velocity cannot exceed the landslide velocity (Pudasaini and Fischer, 2020), $u^{ev} = u-u^b$ is a non-negative quantity.  For convenience, we write (\ref{Eqn_7}) in terms of $u - u^b$:
\begin{equation}
m\frac{du}{dt} + \left (u - u^b\right ) \frac{dm}{dt} = F.                                          
\label{Eqn_8}
\end{equation}
Now, we can compare (\ref{Eqn_8}) with (\ref{Eqn_3}), which is written in the depth-averaged form and for a constant mass density. So, without loss of generality, we can carefully, and consistently set $dm/dt = E$, $m = 1$ (means, per unit mass), $\partial u/\partial t + u \partial u/\partial x = du/dt$, yielding  
\begin{equation}
\frac{du}{dt} 
= g^x - (1-\gamma^m)\alpha^m g^z\mu^m 
-g^z\left [ \left( \left( 1-\gamma^m\right)K + \gamma^m\right)\alpha^m+\left ( 1-\alpha^m \right )\right ]
\frac{\partial h}{\partial x}
-C_{DV}u^2 
+ \left (2u^b -u\right) E\frac{1}{h},                                                              
\label{Eqn_10}
\end{equation}
where, out of $2 u^bE$, one $u^bE$ already exists in the force terms in $F$ that entered as momentum production (Pudasaini and Fischer, 2020), as seen in (\ref{Eqn_2}); 
 however, the other $u^bE$ emerges from the correct handling of the erosion-induced changed inertia.
We can draw an important conclusion from (\ref{Eqn_10}): Since for erosion $E > 0$, whether the erosion related mass flow mobility will be enhanced, reduced or neutralized (remains unaltered) depends exclusively on whether $\left (2u^b -u\right) > 0$, $\left (2u^b -u\right) < 0$, or $\left (2u^b -u\right) = 0$.
 This has been exclusively elaborated in the following sections.
\\[3mm]
Equation (\ref{Eqn_10}) can be cast in different forms. Following Pudasaini and Fischer (2020), we can write $u^b = \lambda^b u$, where $\lambda^b$ is the erosion drift (associated with the erosion velocity). So, (\ref{Eqn_10}) reduces to
\begin{equation}
\displaystyle{
\frac{du}{dt} 
= g^x - (1-\gamma^m)\alpha^m g^z\mu^m 
-g^z\left [ \left( \left( 1-\gamma^m\right)K + \gamma^m\right)\alpha^m+\left ( 1-\alpha^m \right )\right ]
\frac{\partial h}{\partial x} 
- C_{DV}u^2 
+ \left (2\lambda^b -1\right) E\frac{u}{h}.}  
\label{Eqn_11}
\end{equation}
The closure for the erosion drift and its influence in landslide mobility has been presented in Section 5. 
\\[3mm]
For the full and better simulation of the erosive landslide, we must numerically integrate (\ref{Eqn_11}) together with (\ref{Eqn_1}) that includes evolution of both the flow velocity and the flow depth. This will be discussed in Section 9.
Here, we are mainly interested in developing a simple model that can be solved analytically to highlight the main essence of erosion-induced energy (momentum) and the associated mobility of the landslide in terms of its velocity. 
\\[3mm]
Further simplification of (\ref{Eqn_11}) is possible. For simplicity, we can parameterize the landslide (or, the flow) depth $h$, and write (\ref{Eqn_11}) as
\begin{equation}
\frac{du}{dt} = \mathcal{A} - \mathcal{C} u^2 + \left (2\lambda^b -1\right) E\frac{u}{h},       
\label{Eqn_12}
\end{equation}
where, $\displaystyle{ \mathcal{A} = g^x - (1-\gamma^m)\alpha^m g^z \mu^m 
-g^z\left [ \left( \left( 1-\gamma^m\right)K + \gamma^m\right)\alpha^m+\left ( 1-\alpha^m \right )\right ]
\frac{\partial h}{\partial x} }$ takes into account the topography induced downslope component of gravity, the first term; effective basal friction including the buoyancy reduced normal load and lubrication, the second term; and the force due to the free surface pressure gradient of the landslide (including the possible presence of the interstitial fluid),
the third term, which, depending on the negative or positive slope of the landslide, will enhance or reduce the motion (Pudasaini and Hutter, 2007). 
As mentioned earlier, for extensional flows, $K \approx 1$, so $\left [ \left( \left( 1-\gamma^m\right)K + \gamma^m\right)\alpha^m+\left ( 1-\alpha^m \right )\right ]$ reduces to unity.
Moreover, $\mathcal{C} = C_{DV}$ is the viscous drag coefficient.  Equation (\ref{Eqn_12}) can be written in the simple form 
\begin{equation}
\frac{du}{dt} = \mathcal{A} +\mathcal{B} u- \mathcal{C} u^2,     				   
\label{Eqn_13}
\end{equation}
where $\mathcal{B} = \left (2\lambda^b -1\right) E/{h}$. 
Equation (\ref{Eqn_13}) can be solved exactly.
\\[3mm]
One can apply any erosion rate $E$ in the above equations. 
As in Pudasaini and Fischer (2020), we consider the drift factor $\lambda^m$ that is associated with the velocity of the particle in the debris mixture at the lowest level, $u^m$, with the mean velocity of the flow, $u$; the relation $ u^m = \lambda^m u$.
Following the mechanical erosion rate model  by Pudasaini and Fischer (2020):
\begin{equation}
E=   \frac{g\cos\zeta \left[ \left ( 1-\gamma^m\right )
\rho^m\mu^m\alpha^m
                                 -\left ( 1-\gamma^b\right )
\rho^b\mu^b\alpha^b\right] }
{\left( \rho^m\lambda^m\alpha^m - \rho^b\lambda^b\alpha^b \right)}\left(\frac{h}{u}\right),
\label{Eqn_15a}
\end{equation}
(\ref{Eqn_12}) can be written as:  
\begin{equation}
\frac{du}{dt} = \mathcal{A} - \mathcal{C} u^2 + \left (2\lambda^b -1\right) E^P,     
\label{Eqn_14}
\end{equation}
with 
\begin{equation}
E^P =   \frac{g\cos\zeta \left[ \left ( 1-\gamma^m\right )
\rho^m\mu^m\alpha^m
                                 -\left ( 1-\gamma^b\right )
\rho^b\mu^b\alpha^b\right] }
{\left( \rho^m\lambda^m\alpha^m - \rho^b\lambda^b\alpha^b \right)},
\label{Eqn_15}
\end{equation}
where for dry flows and substrate, $\alpha^m$ and $\alpha^b$ are unity, otherwise these must be parameterized or closed.
 Furthermore, as for the sliding mass, the parameters are considered analogously for the erodible basal substrate as indicated by the superscript $^b$.
We call $E^P$ the erosion parameter, which as given by (\ref{Eqn_15}), incorporates many essential physical and mechanical aspects involved in erosion, and explicitly determines the erosion intensity.
The great advantage of (\ref{Eqn_14}) is that the erosion-enhanced flow mobility can now be explicitly evaluated in terms of velocity, as all the quantities (except $u$) on the right hand side of (\ref{Eqn_14}) are measurable, or given. This is the first-ever physics-based model to do so. Thus, it has enormous application potential.
\\[3mm]
{\bf The landslide mobility equation:} 
It is now so convenient that (\ref{Eqn_14}) can be simply written as  
\begin{equation}
\frac{du}{dt} = \left(\mathcal{A}+\mathcal{P}_M\right) - \mathcal{C} u^2,    		     
\label{Eqn_16}
\end{equation}
where $\mathcal{P}_M = \left (2\lambda^b -1\right) E^P$ is the overall mobility parameter (the erosion-induced momentum per unit depth or the force per unit mass) that quantifies the total erosion related enhanced mass flow mobility by amplifying the landslide acceleration. We call (\ref{Eqn_16}) the landslide mobility equation, which can be solved analytically to obtain the landslide velocity with erosion. 
\\[3mm]
Since $u^b = \lambda^b u$, the erosion velocity is associated with the parameter $\lambda^b$. The
form of $E^P$ in (\ref{Eqn_15}) contains no odds. First, in reality, $\lambda^b$ lies in
a close or broader neighborhood of 1/2 that is contained in (0, 1). So, the
legitimate values of $\lambda^b$ is around 1/2.  Second, mechanically,
the erosion velocity is controlled by the net shear stress (applied by the flow minus resisted by the bed material).
 This means, the manner by which $\lambda^b$ changes is controlled by the numerator or the net shear stress.
In other words, in connection to the erosion drift equation
(see below), in total, the higher value of $\lambda^b$ usually corresponds to the higher
mobility parameter $\mathcal{P}_M$.

\section{The State of Energy and Mobility of an Erosive Landslide}

Mobility, perhaps, is the most important aspect in landslide modelling as it is the direct measure of the threat posed by the landslide, and is simply associated with its excessive volume (or, mass), enormous impact energy, the exceptional travel distance and the wide spread inundation  area. Mobility is governed by the state of energy of the landslide and is expressed in terms of the landslide velocity. So, here, we focus on landslide energy budget.
The state of mobility is associated with the sign of $\left (2\lambda^b -1\right)$ and is amplified by the factor $E^P$ in $\mathcal{P}_M =  \left (2\lambda^b -1\right) E^P$ in (\ref{Eqn_16}). We call $ \left (2\lambda^b -1\right)$ the energy generator (or the mobility generator), and write as $ \mathcal{P}_{M_{eg}} =\left (2\lambda^b -1\right)$, the parameter that generates the excess mass flow mobility due to erosion. Mass flow mobility will be enhanced, reduced or remains unchanged depending on whether $ \left (2\lambda^b -1\right) >0$, $ \left (2\lambda^b -1\right) <0$, or $ \left (2\lambda^b -1\right) =0$. As $ E^P $ determines the erosion magnitude, it is of utmost importance to systematically analyze $ \left (2\lambda^b -1\right)$, because this will tell us the state of mobility (associated with the sign), and how the erosion is amplified (its magnitude) that ultimately regulates the strength and consequence of erosion as measured by the landslide velocity. This is how the energy generator changes the game and fully controls the mobility of the erosive landslide.
\\[3mm]
In general, $\lambda^b $ may take any value in the domain (0, 1). However, in solving some engineering and applied problems, we need to physically constrain $\lambda^b $. There can be different possibilities for this, but Pudasaini and Fischer (2020) provide a physical model for $\lambda^b $ by presenting an analytical erosion drift equation:
\begin{equation}
\lambda^m = \left ( 1 + \frac{\rho^b}{\rho^m} \frac{\alpha^b}{\alpha^m}\right ) \lambda^b.	
\label{Eqn_17}
\end{equation}
As for  $\lambda^b $, $\lambda^m$ also takes the values in the domain $(0,1)$. However, in general, as proven by (\ref{Eqn_17}), the velocity of the eroded particle cannot be larger than the velocity of the particle at the flow bottom, we have the constrain $0 <\lambda^b < \lambda^m < 1$. As discussed in Pudasaini and Fischer (2020), the drift equation (\ref{Eqn_17}) is mechanically rich. Following the exclusive consideration of the shallow flow models in the literature (Pudasaini and Hutter, 2007; Le and Pitman, 2009), we may simplify the situation by assuming the plug flow which implies that $\lambda^m \approx 1$. Now, it becomes mechanically very interesting to analyze the landslide mobility with (\ref{Eqn_17}). Below, we consider three special situations, with respect to the inertial number,
$N_i = {\rho^b\alpha^b}/{\rho^m\alpha^m}$.

\subsection{The erodible bed substrate is inertially weaker: Enhanced energy and mobility}
 
The inertia of the material in the bed can be lower than the inertia of the material in the flow: $\rho^b\alpha^b < \rho^m\alpha^m $. Then, we obtain $\lambda^b > 1/2 \lambda^m $, which implies that $\lambda^b \in \left ( 1/2, 1\right )$, and $\left ( 2 \lambda^b -1\right ) \in (0,1)$.  In other words, this is the situation in which the change in inertia of the landslide in incorporating the inertially (mechanically) weaker material is less than its change in inertia if it would have incorporated inertially  equally stronger material. This suggests that the erosion-induced gained momentum or energy results in the enhanced-mobility of the erosive landslide. 

\subsection{The erodible bed substrate is inertially stronger: Reduced energy and mobility}

The inertia of the bed material can be higher than inertia of the material in the flow: $\rho^b\alpha^b > \rho^m\alpha^m $. This implies $\lambda^b < 1/2 \lambda^m $, and $\lambda^b \in \left (0, 1/2\right )$. Thus, $\left ( 2 \lambda^b -1\right ) \in (-1,0)$.  So, for this, the change in inertia of the landslide in incorporating the inertially stronger material is higher than its change in inertia if it would have incorporated inertially equally stronger material. This implies the erosion-induced momentum loss or energy loss, and results in the reduced mobility even for the erosive landslide.

\subsection{The erodible bed substrate is inertially neutral: No change in energy and mobility}

 For this, the inertia of the bed material is equal to the inertia of the material in the flow: $\rho^b\alpha^b = \rho^m\alpha^m $. Thus, we obtain $\lambda^b = 1/2 \lambda^m $, which implies that $\lambda^b = 1/2$, and $2 \lambda^b -1 = 0$. In this situation, there is no gain or loss of momentum or energy, and thus, the landslide mobility remains unchanged even for the erosive landslide. This is a very special situation, however, less likely to occur in nature. 

 \subsection{A single frame describing enhanced energy and mobility}
 
The conditions in Section 5.1 - Section 5.3 can be unified into a single frame: 
$\left(2\lambda^b-1\right) \in (-1, 1) = \left( -1, 0\right)\cup \left\{0\right\}\cup\left(0, 1\right)$ for
$\lambda^b \in (0, 1) = \left( 0, 1/2\right)\cup \left\{1/2\right\}\cup\left( 1/2, 1\right)$,
covering the whole spectrum of momentum or energy loss (Section 5.2), or equilibrium (Section 5.3), or gain (Section 5.1), that result in reduced, neutral or enhanced landslide mobility. With the knowledge of the energy generator $ \mathcal P_{M_{eg}} =\left (2\lambda^b -1\right)$, the involved net energy in landslide erosion can now be quantified from the mobility parameter $\mathcal{P}_M$. Such an explicit and fully mechanical description of the state of energy (or, momentum) and the associated mobility of an erosive landslide is seminal.  
\\[3mm]
As $\lambda^b$ is related to $1/2\lambda^m$ and $\lambda^m \approx 1$,
following the analysis in Pudasaini and Fischer (2020), and the above
discussion, technically suitable natural domain of $\lambda^b$ is: $\left(-\Delta
\lambda +1/2, 1/2 + \Delta \lambda\right)$, where $\Delta \lambda$ is a small
positive number, say, typically 1/4, such that the value of $\lambda^b$ is
always contained in (0, 1). The drift factor $\lambda^b$ is more likely to approach 1 rather
than to 0 indicating the energy gain than the energy loss in erosion. 

\section{The Outstanding Role of Erosion Velocity in Mass Flow Mobility}

In many of the previous erosion models the velocity of the eroded mass has been set to zero, or it does not appear at all. For the first time, Pudasaini and Fischer (2020) rigorously proved with a mechanical erosion model that setting the erosion velocity to zero is physically incorrect. In this line, the above analysis clearly expands our understanding of erosion related phenomena and shows that, whether the erosive landslide will gain or lose (or, remain unchanged) energy, or in other words, whether it will have enhanced or reduced (or, neutral) mobility as compared to the non-erosive one, primarily depends on the velocity of the eroded mass $u^b$. In technical terms, it depends on the value of the drift factor $\lambda^b$ explaining how big is the erosion velocity with respect to the flow velocity. Erosion velocity closer to the flow velocity results in the most mobile flow. Because, in this situation, the momentum production $\left(u^b E\right)$ due to the reduced friction in erosional situation overtakes the momentum loss due to the increased inertia $\left (u-u^b\right)E$. This is how most probably it happens in nature for erosive landslides. As $u^b \to u$, the increase in inertia associated with the entrainment tends to vanish. Then, the flow attains the highest gain in energy resulting in the highest mobility, as measured by the gained or produced momentum $\left(u^b E\right)$ of the flow due to erosion. 
This analysis clearly reveals the fact, that paired with the momentum production and correct handling of the change of inertia in describing the erosion related energy, the erosion velocity plays a key and outstanding role in appropriately determining the energy budget and mobility of an erosive landslide. 
\\[3mm]
Moreover, if the erosion velocity is less than one half of the flow velocity then, the landslide loses energy. This results in the reduced mobility. The highest energy is consumed in the erosion if the erosion velocity is much smaller than the flow velocity, this means when the erosion velocity is almost zero. In this situation, $(-uE)$ is the reduced momentum, which is produced by the increased inertia due to the entrainment when the entrained mass enters the landslide with zero velocity. Physically, this is impossible as proved by Pudasaini and Fischer (2020), as $\lambda^b \neq 0$, however, this refers to the situation in many previous erosion models (Perla et al., 1980; Fraccarolla and Capart, 2002; McDougall and Hungr, 2005; Christen et al., 2010; Iverson, 2012; Frank et al., 2017).
\\[3mm]
Interestingly, the erosion will not change the energy status, and thus the mobility, of the landslide if the erosion velocity is one half of the flow velocity.  Such a special situation has also been mentioned in Le and Pitman (2009) and Pudasaini and Fischer (2020), which, however, is very restricted, and less likely to happen in nature. So, the present paradigm further enhances the mechanical erosion model by Pudasaini and Fischer (2020) and offers a complete model for landslide erosion.  

\section{The Erosion-, Entrainment-, and Energy-Velocity:\\ New Concepts with Mechanics}

Here, we formally introduce three important concepts with their underlying mechanics. These are: (i) the erosion-velocity, $u^e=u^b$, (ii) the entrainment-velocity, $u^{ev} = u-u^b$, and (iii) the energy-velocity, $u^{env} = u^b - \left (u - u^b\right)$. While the erosion velocity was first introduced by Pudasaini and Fischer (2020), the entrainment velocity and the energy velocity are completely new concepts.
 In fact, $u^{e}$, $u^{ev}$ and $u^{env}$ already appear in previous sections. $u^{env} = u^b - \left (u - u^b\right) = 2u^b - u$ plays exclusively unique and dominant role in formulating the mobility equation and in determining the state of energy, and thus mobility. As it is clear from (\ref{Eqn_10}), $u^{env}$ is the total (net) momentum production with contribution $u^b$ emerging from the reduced friction and $-\left (u - u^b\right) $ from the changed (reduced) inertia. As $u^{env}$ has the dimension of velocity and generates the erosion-induced excess energy that enhances the mobility, this is called the energy-velocity. Thus, the energy-velocity provides the universal picture of the erosion-induced mobility.  

\subsection{The landslide-propulsion and erosion-thrust}

 With these definitions, their mechanics and the discussions in the previous sections, we draw a central conclusion: the landslide gains energy, and thus enhances its mobility if the energy-velocity is positive, specifically, if the erosion velocity is greater
than the entrainment velocity, i.e., $u^e > u^{ev}$. We call this phenomena the landslide-propulsion, emerging from the net momentum production, that provides the erosion-thrust to the landslide. This means, if the
erosion velocity is greater than one half of the flow velocity, i.e.,
$u^b > u/2$, the mobility is enhanced. This is equivalent to the condition $\lambda^b > 1/2$. In
other words, the landslide gains energy to enhance its mobility if the
eroded material is easily entrainable with the velocity lower than the
erosion velocity. These are quite natural phenomena, but revealed here for
the first time. 

\subsection{Distinction between erosion and entrainmnet}

The existing literature could not distinguish between the erosion and entrainment as these terms are used interchangeably. However, here, we have made it very clear with the mechanical expressions, that the erosion and entrainment are essentially different phenomena. Erosion is a process by which the bed material is mobilized by the flow with the velocity $u^e = u^b$, while entrainment is intrinsically another process by which the eroded material is incorporated (entrained) and taken along with by the flow with the velocity $u^{ev} = u-u^b$. This fundamentally enhances our understanding of basic, but different processes in erosion related phenomena in landslide by clearly defining, and distinguishing the mechanisms of erosion and entrainment. These are important novel aspects.

\subsection{Delineating different energy regimes}

The energy velocity, $u^{env} = u^b - \left (u – u^b\right) = u^e - u^{ev}$, which constitutes the state of energy or the energy budget, is the erosion
velocity in excess to the entrainment velocity. $u^{env}$ clearly delineates
the three energy regimes associated with the erosive landslide: 
\begin{itemize}
\item The landslide gains energy in erosion if the energy-velocity is positive, $u^{env} > 0$. 

\item  The landslide loses
energy even in erosion if the energy-velocity is negative, $u^{env} < 0$. 

 \item  The landslide energy
remains unchanged if the energy-velocity is zero, $u^{env} = 0$. 
\end{itemize}
In terms of $\lambda^b$, these regimes
correspond to $\lambda^b > 1/2$, $\lambda^b < 1/2$, and $\lambda^b = 0$, respectively.
 So, the energy, and thus the mobility, of an erosive landslide is fully controlled by the erosion velocity. This signifies the prime role of erosion velocity in correctly determining the state of erosive landslide.

\section{Analytical Solution to the Landslide Mobility}

\subsection{Exact analytical solution}

The landslide mobility equation (\ref{Eqn_16}) can be solved analytically to explicitly obtain the landslide velocity. Exact analytical solutions to simplified cases of nonlinear debris avalanche model equations are necessary to calibrate numerical simulations (Pudasaini, 2016). These problem-specific solutions provide important insights into the full behavior of the system. A physically meaningful exact solution explains the true and entire nature of the problem associated with the model equation (Pudasaini, 2011; Faug, 2015; Pudasaini et al., 2018; Pilvar et al., 2019). 
 However, numerical solutions are always subject to questions as such solutions are based on some assumptions and applied approximations that may not follow the laws of nature. So, the physically relevant exact solutions are superior over the numerical simulations (Pudasaini and Krautblatter, 2021). 
\\[3mm]
The models (\ref{Eqn_16}) can be solved either in Eulerian form with the left hand side written as $\partial u/\partial t + u \partial u/\partial x$, or in the Lagrangian form written as $du/dt$.  
Since here we aim to explicitly quantify the effect of erosion in landslide velocity, for simplicity, we consider (\ref{Eqn_16}) in Lagrangian form and obtain the exact landslide velocity. However, we mention, that the exact solution of (\ref{Eqn_16}) can also be obtained in Eulerian form, but is very demanding mathematically. Pudasaini and Krautblatter (2021) have presented various exact analytical solutions for landslide velocity, however, without considering the erosion effects. Here, we focus on the velocity of an erosive landslide.
\\[3mm]
Now, we present the dynamics of the landslide mobility model. The model (\ref{Eqn_16}) is a first-order non-linear ordinary differential equation which possesses an exact analytical solution for the time-evolution of the landslide velocity in the form of the tangent hyperbolic function:
\begin{equation}
u(t) = \sqrt{\frac{\mathcal A +\mathcal{P}_M}{\mathcal C}}\tanh\left[ 
\sqrt{\left(\mathcal A + \mathcal{P}_M\right) \mathcal{C}}\,\left(t-t_i\right) + \tanh^{-1}\left (\sqrt{\frac{\mathcal C}{\mathcal A +\mathcal{P}_M}}\,\,u_i\right)
\right],
\label{Eqn_19}
\end{equation}
where, $u_i$ is the initial (or, boundary) condition at a given time $t = t_i$. 

\subsection{Steady $-$ state velocity and its importance}

For sufficiently large time (equivalently, sufficiently long distance), (\ref{Eqn_19}) can be represented as 
\begin{equation}
\lim_{t \to \infty} u(t) = \sqrt{\frac{\mathcal A +\mathcal{P}_M}{\mathcal C}},
\label{Eqn_20}
\end{equation}
which is the steady-state (uniform) velocity of the landslide, that is determined by the applied forces $\mathcal A$ and $\mathcal C$, and the erosion-induced mobility parameter $\mathcal{P}_M$. 
This particular solution could already be obtained from (\ref{Eqn_16}) by assuming the steady-state condition, $0 = {\mathcal A +\mathcal{P}_M} -{\mathcal C} u^2$.
The explicit time-independent form of the velocity in (\ref{Eqn_20}) is important in quickly solving technical and engineering problems. We call it the representative (steady-state) landslide mobility velocity, $u^{lm}_s$, and write
\begin{equation}
u^{lm}_s = \sqrt{\frac{\mathcal A +\mathcal{P}_M}{\mathcal C}}.
\label{Eqn_21}
\end{equation}
Although it is simple as it appears, (\ref{Eqn_21}) includes many of the dominant and essential physical aspects of the material and the flow as carried by the definitions of $\mathcal A, \mathcal{P}_M$ and $\mathcal C$. The involved parameters can be estimated from their definitions, and depending on the situation, can have wide range of values. 
\\[3mm]
To quantify $u$ in (\ref{Eqn_19}) and (\ref{Eqn_21}), we consider the often used and physically plausible parameter values with appropriate dimensions (Mergili et al., 2020a, 2020b; Pudasaini and Fischer, 2020) as follows: for sliding mass: 
$\delta^m = 40^\circ, \gamma^m = \rho^m_f/\rho^m_s = 1100/2900, \alpha^m = 0.75$; for the erodible basal substrate: $\delta^b = 10^\circ, \gamma^b = \rho^b_f/\rho^b_s = 1000/2000, \alpha^b = 0.5, \lambda^b = 0.69$; where $\lambda^b$ is computed from these parameters. Furthermore, we consider a slope inclined at an angle $\zeta = 45^\circ$. 
With these, we obtain the typical values of the model parameters as: $\mathcal {A} = 4.2271$, $\mathcal{P}_M = 1.7988$; and utilize $\mathcal C = 0.0014$. This results in some representative velocities of fast moving landslides: $u^{lm}_s = 55$ ms$^{-1}$ without erosion and  $u^{lm}_s = 65.6$ ms$^{-1}$ with erosion, which already shows significant difference between these velocities. However, based on the parameter values, the relative difference in velocities, with and without erosion, can be even higher as $\mathcal{P}_M$ might possibly be higher than $\mathcal{A}$. Here, we have just presented a possible scenario. 
These velocities are quite reasonable for fast to rapid landslides and debris avalanches and correspond to several natural events (Highland and Bobrowsky, 2008). Simulation results show that the front of the 2017 Piz-Chengalo Bondo landslide (Switzerland) moved with more than 25 ms$^{-1}$ already after 20 s of the rock avalanche release (Mergili et al., 2020b), and later it moved at about 50 ms$^{-1}$ (Walter et al., 2020). The 1970 rock-ice avalanche event in Nevado Huascaran (Peru) reached a mean velocity of 50 - 85 ms$^{-1}$ at about 20 s, but the maximum velocity in the initial stage of the movement reached as high as 125 ms$^{-1}$ (Erismann and Abele, 2001; Evans et al., 2009; Mergili et al. 2018). The 2002 Kolka glacier rock-ice avalanche in the Russian Kaucasus accelerated with the velocity of about 60 - 80 ms$^{-1}$, but also attained the velocity as high as 100 ms$^{-1}$, mainly after the incipient motion (Huggel et al., 2005; Evans et al., 2009). All these events were substantially to highly erosive.
 By properly selecting the model parameters such exceptionally high velocities as inferred from the field can be obtained from the new model. 

\subsection{Time evolution of landslide velocity with erosion}

\begin{figure}
\begin{center}
  \includegraphics[width=\linewidth]{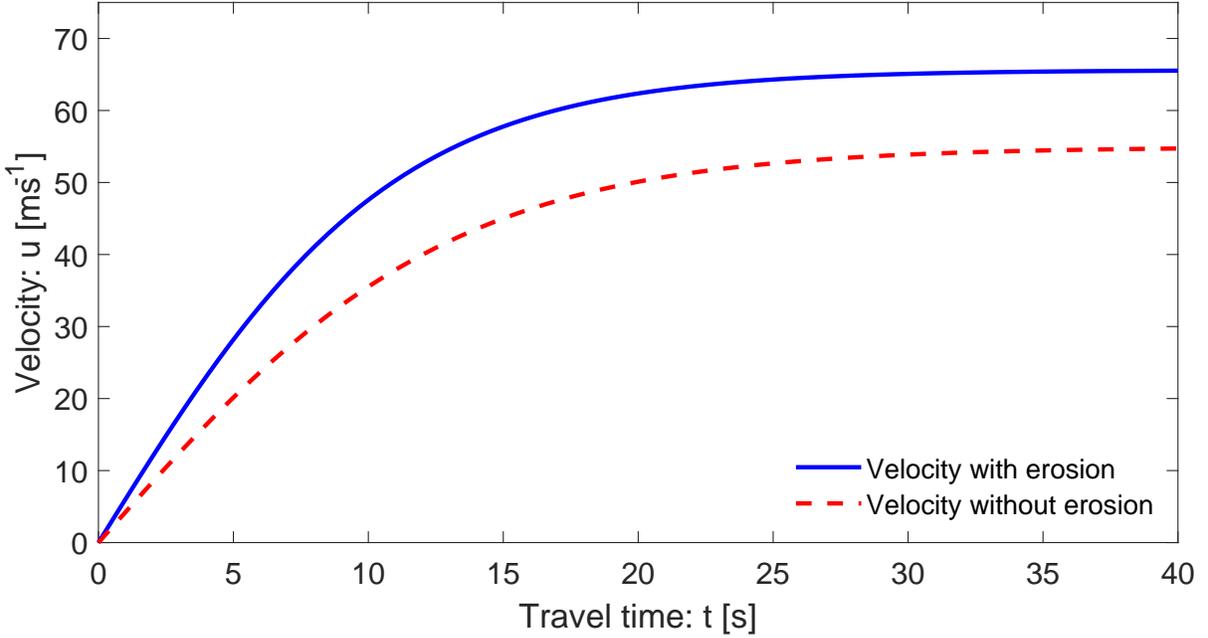}
  \end{center}
  \caption[]{Time evolution of the landslide velocity with and without erosion given by (\ref{Eqn_19}). Erosion enhances the landslide velocity and thus its mobility. With erosion, the steady state velocity is higher and is reached earlier than the same without the erosion. This is due to the erosion-induced gain in momentum that increases the instantaneous velocity for which the drag takes shorter time to bring the motion to the steady-state, but with higher value.}
  \label{Fig_1}
\end{figure}

The full time evolution of the landslide velocity with erosion given by (\ref{Eqn_19}) has been shown in Fig. \ref{Fig_1} with $u_i = 0$ at $t_i = 0$. The flow dynamics is controlled by the competition (interaction) between the overall (net) driving and the resisting forces, ${\mathcal A +\mathcal{P}_M}$ and ${\mathcal C}u^2$, respectively. Importantly, if the initial velocity is less than the steady-state velocity, i.e., $u_i < u^{lm}_s $, then after its inception, the landslide accelerates (rapidly or slowly, depends on the magnitude of $u^{lm}_s -u_i$) because ${\mathcal A +\mathcal{P}_M}$ dominates ${\mathcal C}u^2$. Example includes the situation when the landslide is initially triggered with zero velocity, e.g., due to the slope failure from its static condition. However, in long time, as ${\mathcal C}u^2$ balances ${\mathcal A +\mathcal{P}_M}$, $u(t)$ asymptotically approaches, from below, the steady-state velocity, $u^{lm}_s$.
This is the situation presented in Fig. \ref{Fig_1}.
On the other hand, if the initial velocity is higher than the steady-state velocity, i.e., $u_i > u^{lm}_s $, then, after its triggering, the landslide decelerates (rapidly or slowly, depends on the magnitude of $u_i - u^{lm}_s$) because ${\mathcal C}u^2$ dominates ${\mathcal A +\mathcal{P}_M}$. The landslide triggered by strong seismic shacking is an example for this. Nevertheless, in long time, as ${\mathcal A +\mathcal{P}_M}$ tends to neutralize ${\mathcal C}u^2$, $u(t)$ asymptotically approaches, from above, the steady-state velocity, $u^{lm}_s$. 
Technically, $u^{lm}_s$ provides an important information of landslide velocity with erosion for landslide engineers and practitioners. 
Equations (\ref{Eqn_19}) and (\ref{Eqn_21}) clearly indicate that the higher the value of the mobility parameter $\mathcal{P}_M$ the earlier the landslide reaches its steady-state with substantially higher velocity. This is quite natural, because as erosion enhances the velocity, it takes relatively shorter time for the drag to control the acceleration of the landslide. In other words, this also proves that erosion enhances mobility for the positive values of the mobility parameter $\mathcal{P}_M$. 

\subsection{Quantifying the importance of erosion} 

\begin{figure}
\begin{center}
  \includegraphics[width=\linewidth]{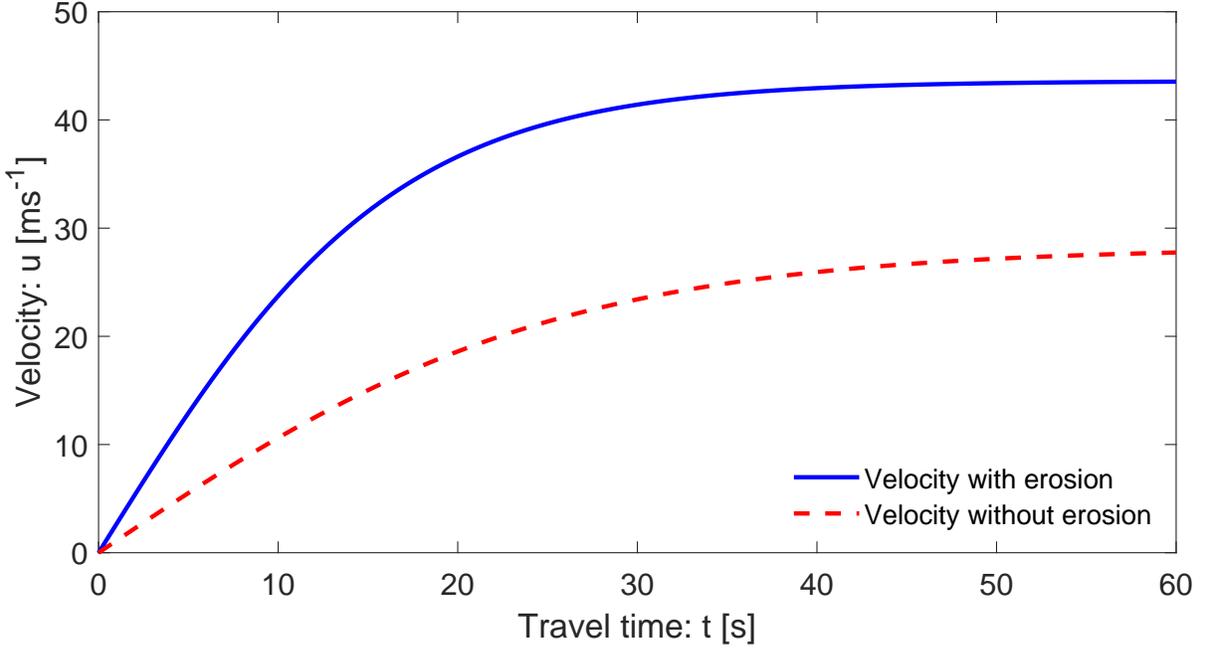}
  \end{center}
  \caption[]{Time evolution of velocity with and without erosion for dry landslide given by (\ref{Eqn_19}). The landslide velocity, and thus its mobility, is largely enhanced by erosion.}
  \label{Fig_11}
\end{figure}

Figure \ref{Fig_1} shows that, around $t = 15$ s, the velocities with and without erosion take values of about 57 and 44~ms$^{-1}$, respectively, with the maximum difference of $13$ ms$^{-1}$. And, in long time, the corresponding steady-state velocities are $65.6$ ms$^{-1}$ and $55$ ms$^{-1}$. As the dynamic pressure is proportional to the square of the velocity, with respect to the steady-state velocities, the dynamic pressure with erosion is about 42\% higher than the same without erosion. 
However, with respect to the maximum difference in the velocities at $t = 15$ s, the dynamic pressure with erosion is even 68\% higher than the same without erosion. 
Crucially, these contrasts in velocities result in completely different run-out and deposition scenarios. This clearly manifests the importance of the correct inclusion of the erosion in modelling the landslide dynamics and run-out. 
\\[3mm]
If we consider both the landslide and the basal substrate consisting of only solid particles and neglect all the fluid related parameters (forces), we need to set $\alpha^m = 1, \alpha^b = 1, \gamma^m = 0, \gamma^b = 0$. Then, the velocities with and without erosion would be much smaller, and attain the steady-state values of $43.56$ ms$^{-1}$ and $28.23$ ms$^{-1}$, respectively. So, the steady-state is reached much later in time. However, the relative difference is 15.23 ms$^{-1}$, which is higher than before. This is because of the strongly reduced value of $\mathcal A$, but $\mathcal{P}_M$ decreases only slightly (to 1.1 and 1.5, respectively). 
The results are presented in Fig. \ref{Fig_11}. Yet, the maximum difference in velocities with and without erosion is about 18.30 ms$^{-1}$ (= 39.4 ms$^{-1}$ $-$ 21.1 ms$^{-1}$) at around $t = 25$ s. So, at this point, the dynamic pressure with erosion is about 2.5 times higher than the same without erosion, which is a huge contrast. 

\subsection{Velocity as a function of travel distance}

\begin{figure}
\begin{center}
\includegraphics[width=\linewidth]{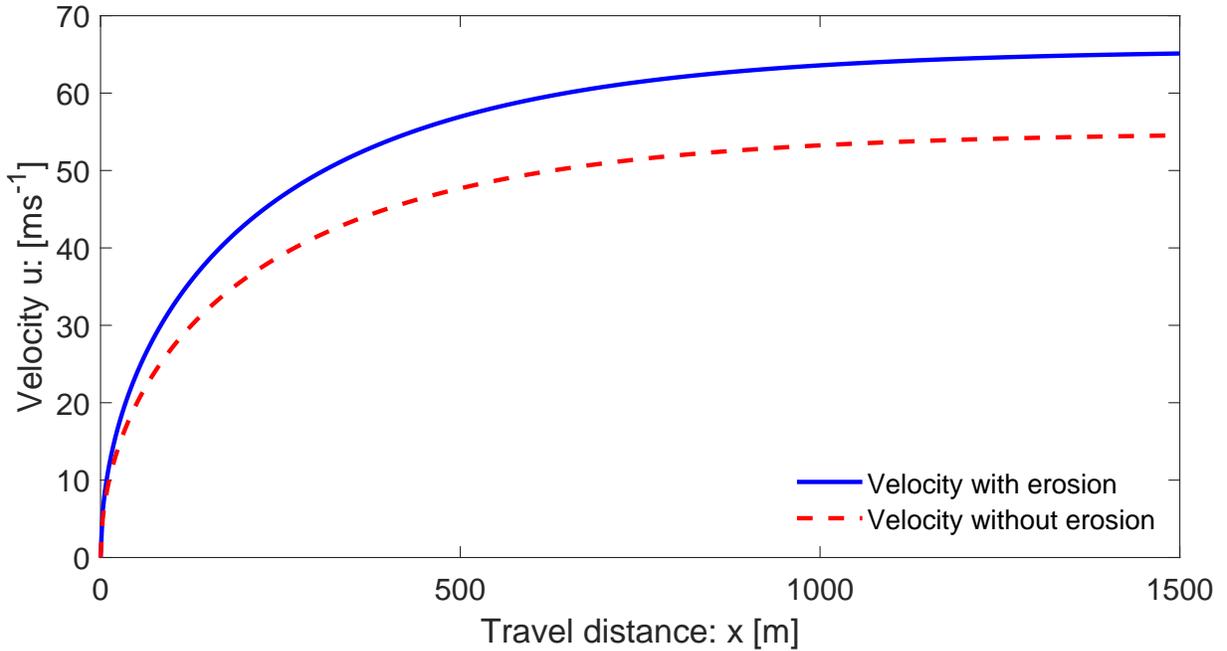}
  \end{center}
  \caption[]{Evolution of the landslide velocity as a function of travel distance with and without erosion given by (\ref{Eqn_16_b}). Erosion enhances the landslide velocity and thus its mobility.}
  \label{Fig_2}
\end{figure}

For a mass point motion, we may write: 
\begin{equation}
\displaystyle{\frac{du}{dt} = \frac{du}{dx}\frac{dx}{dt}  = u \frac{du}{dx}}. 
\label{Eqn_16_0}
\end{equation}
Then, (\ref{Eqn_16}) takes the form
\begin{equation}
u\frac{du}{dx} = \left(\mathcal{A}+\mathcal{P}_M\right) - \mathcal{C} u^2,    		     
\label{Eqn_16_a}
\end{equation}
which can be solved analytically to obtain exact solution for the landslide velocity as a function of travel distance: 
\begin{equation}
 u(x) = \sqrt{\frac{\mathcal A + \mathcal P_M}{\mathcal C}\left[ 1 - \left ( 1- \frac{\mathcal C}{\mathcal A + \mathcal P_M}u_i^2\right)\frac{1}{\exp(2\mathcal C(x-x_i))} \right]}\,,
\label{Eqn_16_b}
\end{equation}
where, $u_i$ is the initial velocity at $x_i$. The results have been presented in Fig. \ref{Fig_2}, where, both velocities have the same limiting values as in Fig. \ref{Fig_1}, otherwise their behaviors are quite different. In space, the velocity shows a hyper increase after the incipient motion. However, the time evolution of velocity is first slow (almost linear) then fast, and finally attains the steady-state, the common limiting value for both the solutions (\ref{Eqn_19}) and (\ref{Eqn_16_b}). These results indicate that, in any situations (Fig. \ref{Fig_1} - Fig. \ref{Fig_2}), the differences in the landslide velocities with and without erosion are huge. This demonstrates the control of erosion over the landslide mobility.

\subsection{The mobility scaling and erosion number}

By considering the simple initial condition $u_i = 0$ at $x_i = 0$, the structure of solution (\ref{Eqn_16_b}) clearly indicates that, there exists a unique number $\mathcal S_M$:
\begin{equation}
 \displaystyle{\mathcal S_M = \sqrt{1 + \frac{\mathcal P_M}{\mathcal A}}}\,, 
\label{Eqn_16_c}
\end{equation}
such that 
\begin{equation}
 u(x) = \mathcal S_M\,u_{n_{er}}(x), \,\,\,\, u_{n_{er}}(x) = \sqrt{\frac{\mathcal A}{\mathcal C}\left[ 1 -\frac{1}{\exp(2\mathcal C x)} \right]}\,,
\label{Eqn_16_d}
\end{equation}
where, $u_{n_{er}}$ is the landslide velocity without erosion. We call $\mathcal S_M$ the mobility scaling. Both mechanically and technically, $\mathcal S_M$ has a great significance. First, it is simple, and exclusively depends on all the measurable physical and mechanical parameters of the landslide, the net driving force ${\mathcal A}$ and the mobility parameter ${\mathcal P_M}$. Second, it is a novel dimensionless number that scales the landslide mobility through velocity. Third, with the knowledge of the mobility parameter ${\mathcal P_M}$, the practitioners can recover the velocity of an erosive landslide from  (\ref{Eqn_16_d}), even previously not knowing the velocity with erosion. This is a special property of the solution (\ref{Eqn_16_b}). This idea can equally be applied for general simulation results. Fourth, $\mathcal S_M$ depends non-linearly on ${\mathcal P_M}$. As discussed in Section 5 and Section 7, ${\mathcal P_M > 0}$, ${\mathcal P_M = 0}$, or ${\mathcal P_M < 0}$ delineate the enhanced, neutralized, or reduced mobility regimes, so the range of $\mathcal S_M$ should be understood accordingly. Hence, for ${\mathcal P_M = 0}$, $\mathcal S_M = 1$ degenerates to the landslide without erosion, while $\mathcal S_M > 1$ for positive value of ${\mathcal P_M}$ corresponds to the erosion-enhanced mobility. However, $\mathcal S_M < 1$ in the negative ${\mathcal P_M}$ domain is that for reduced mobility. While ${\mathcal P_M}$ delivers the overall mobility as the additional force induced by erosion in the dynamical system (\ref{Eqn_16_b}), the mobility scaling $\mathcal S_M$ provides us with the direct and explicit measure of mobility by contrasting the landslide dynamics without erosion from that with erosion. As $\mathcal S_M$ exactly quantifies the contribution of erosion in landslide mobility, technically, this is the most attractive and pleasant feature of the mobility scaling. 
\begin{figure}
\begin{center}
\includegraphics[width=\linewidth]{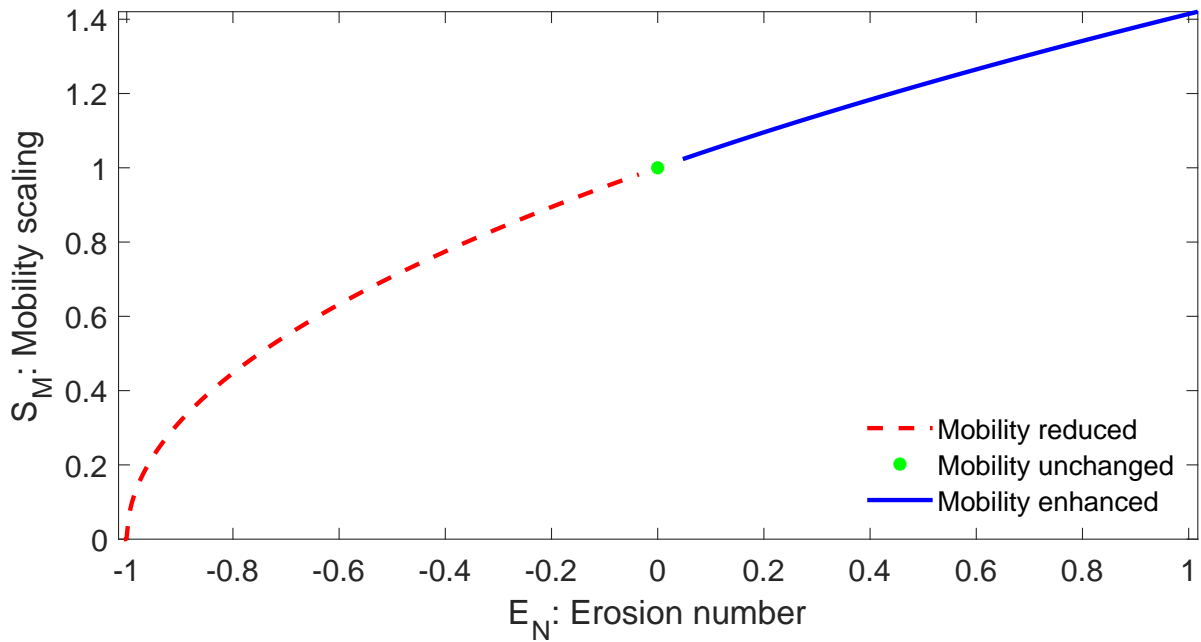}
  \end{center}
  \caption[]{Non-linear dependency of the mobility scaling $\mathcal S_M$ on the erosion number ${\mathcal E_N}$ given by (\ref{Eqn_16_c}) and (\ref{Eqn_16_e}). Three distinct mobility regimes are indicated.}
  \label{Fig_2_S_M}
\end{figure}
\\[3mm]
In the definition of $\mathcal S_M$ in (\ref{Eqn_16_c}), the ratio 
\begin{equation}
{\mathcal E_N = \frac{\mathcal P_M}{\mathcal A}},
\label{Eqn_16_e}
\end{equation}
plays a central role. We call $\mathcal E_N$ the erosion number. The erosion number $\mathcal E_N$ is the second novel dimensionless number presented here as a ratio between the erosion-induced force ${\mathcal P_M}$ (also called the mobility parameter) and the net driving force ${\mathcal A}$ (per unit mass). Depending on whether ${\mathcal P_M}$ is positive, zero or, negative, $\mathcal E_N$ can be positive, zero or, negative implying the enhanced, unaltered or, reduced mobility. In connection to the definition of the mobility scaling $\mathcal S_M$, the possible negative value of $\mathcal E_N$ is the structural requirement, however, is not an odd. We may also call $\mathcal E_N$ the erosion-mobility number. The dependency of the mobility scaling as a function of the erosion number has been shown in Fig. \ref{Fig_2_S_M}. The parameter values for the results presented in Fig. \ref{Fig_2_S_M} are: $\mathcal A = 4.2271, \mathcal P_M = 1.7988, \mathcal E_N = 0.4258, \mathcal S_M = 1.1941$, respectively.

\subsection{Negative overall net driving force}

We note that in situations when $\displaystyle{\frac{\mathcal P_M}{\mathcal A} > -1}$ or ${\mathcal A} + {\mathcal P_M} < 0$, we must re-derive the solutions to replace (\ref{Eqn_19}) and (\ref{Eqn_16_b}), as the new solutions would be structurally and dynamically different because of the changed interactions between the associated system forces. However, we do not elaborate in this aspect here.

\subsection{Reduction to classical Voellmy mass point model}

We note that, in the structure, the model (\ref{Eqn_16}) or, (\ref{Eqn_16_a}), and its solution (\ref{Eqn_19}) or, (\ref{Eqn_16_b})  exists in literature (Pudasaini and Hutter, 2007) and is classically called Voellmy's mass point model (Voellmy, 1955; Gruber, 1989), or Voellmy-Salm model (Salm, 1966; McClung et al., 1983). Perla et al. (1980) also called (\ref{Eqn_16}) the governing equation for the center of mass velocity. However, $(1-\gamma^m)$, $\alpha^m$, and the term associated with $\partial h/\partial x$, and erosion are the new contributions and were not included in the Voellmy model, and $K = 1$ therein, while in our consideration of $\mathcal A$, $K$ can be chosen appropriately. Thus, the Voellmy model corresponds to the substantially reduced form of $\mathcal A$, with $\mathcal A = g^x-g^z \mu$, and $\mathcal P_M = 0$. 

\section{Complete Set of Dynamical Landslide Equations with Erosion}

We have now the proper understanding of the structure of the inertial part
of the momentum equation for
 landslide with erosion in Section 4.
For the full and better simulation of the erosive landslide problem, we
must numerically integrate the mass and momentum balance equations
that include the evolution of both
the state variables, the flow velocity and the flow depth.
We can now formulate the full set of dynamical equations with erosion either in the
non-conservative from, or in conservative form.
However, in general, it is better to use the non-singular mechanical erosion rate as derived
in Pudasaini and Fischer (2020):
\begin{equation}
\displaystyle {E =   \frac{\sqrt{g \cos\zeta h\left[ {\left ( 1-\gamma^m\right )
\rho^m\mu^m\alpha^m
                                 -\left ( 1-\gamma^b\right )
\rho^b\mu^b\alpha^b}\right] }}
{\sqrt{\nu\left( \rho^m\lambda^m\alpha^m - \rho^b\lambda^b\alpha^b \right)}}\,},
\label{Eqn_15a_CON}
\end{equation}
where, $\nu$ connects the shear velocity of the system with the flow velocity. 
 Usually, the shear velocity is about 5\% to 10\% of the mean flow velocity. 
So, for simplicity, 
we can take a suitable value of $\nu$ in the domain (100, 400). Otherwise, we follow
Pudasaini and Fischer (2020) for an analytical closure relation for $\nu$.

\subsection{Non-conservative formulation}

The momentum balance equation (\ref{Eqn_11}),
together with the mass balance (\ref{Eqn_1}),
constitutes the set of full dynamical model equations for erosive landslide in
non-conservative form
\begin{equation}
\frac{\partial h}{\partial t} +  \frac{\partial }{\partial x}\left ( hu\right ) = E,
\label{Eqn_1_NON_C}
\end{equation}
\begin{equation}
\displaystyle{
\frac{du}{dt} 
= g^x - (1-\gamma^m)\alpha^m g^z\mu^m 
-g^z\left [ \left( \left( 1-\gamma^m\right)K + \gamma^m\right)\alpha^m+\left ( 1-\alpha^m \right )\right ]
\frac{\partial h}{\partial x} 
- C_{DV}u^2 
+ \left (2\lambda^b -1\right) E^p,}  
\label{Eqn_11NON_C}
\end{equation}
where, $E^p = E u/h$.
These equations include evolution of both the flow velocity and the flow
depth.

\subsection{The conservative formulation}

Due to the possible rapid spatial and temporal changes of the flow variables, in practice, the
mass flow model equations are often exclusively solved
in conservative form (Tai et al., 2002; Pudasaini and Hutter, 2007; Pudasaini and Mergili, 2019; Mergili et al., 2020a,b). So, we need to formulate the balance equations in conservative form.
First, we note that the third term on
the left hand side of (2), i.e.,
$\displaystyle{\frac{\partial}{\partial x}\left [ \left ( 1-
\gamma^m\right)\alpha^m g^z K{h^2}/{2}\right]}$ emerges from the stress (particularly known as the hydraulic pressure gradient),
 and thus not a part of the inertia of the system. Classically, this term is
organized that way only for the convenience and
for analytical or numerical reasons (Pudasaini and Krautblatter, 2014; Kattel et al., 2018; Kafle et al., 2019).
We have seen that, for the erosive landslide, the main essence lies in the use of the correct velocity
(the entrainment velocity) in transporting the newly entrained mass (the rate of change of
landslide mass) that appears in the inertial part of the momentum balance equation.
With this, the non-conservative momentum equation (\ref{Eqn_3}) can be structurally
brought back to the conservative form (\ref{Eqn_2}).
This is achieved by legitimately and consistently re-writing (\ref{Eqn_3}) as:
\begin{eqnarray}
&&h\left[\frac{\partial u}{\partial t} + u \frac{\partial u}{\partial x}\right]
+ \left(u-u^b\right)\left[\frac{\partial h}{\partial t} + \frac{\partial}{\partial x}\left ( hu\right )\right]\nonumber\\
&& = h\left[g^x - (1-\gamma^m)\alpha^m g^z\mu^m 
 -g^z\left \{ \left( \left( 1-\gamma^m\right)K + \gamma^m\right)\alpha^m+\left ( 1-\alpha^m \right )\right \}
 \frac{\partial h}{\partial x}
 -C_{DV}u^2 \right]+ u^b E.  
 \label{Eqn_3_CON_A}
\end{eqnarray}
The physical reason for the appearance of $\left( u - u^b\right)$, on the left hand side, for the erosional landslide has been explained in Section 3.
With the help of (\ref{Eqn_1_NON_C}), this can be written as
\begin{eqnarray}
&&\frac{\partial}{\partial t}\left( hu \right)+ \frac{\partial}{\partial x} \left [ hu^2 + \left ( 1- \gamma^m\right)\alpha^m g^z K\frac{h^2}{2}\right]\nonumber\\ 
&&= h \left[g^x - \left ( 1-\gamma^m \right ) \alpha^m g^z\mu^m
-\left\{ 1-\left( 1-\gamma^m\right)\alpha^m\right\} g^z\frac{\partial h}{\partial x} 
- C_{DV} u^2\right] + 2u^b E.
\label{Eqn_2_CON_B}
\end{eqnarray}
This completes the process of deriving the full set of dynamical equations (mass and momentum balances) for erosional landslide in conservative form.
\\[3mm]
As it is clear from the derivation, the momentum balance equation (\ref{Eqn_2_CON_B}) correctly includes the erosion-induced change in inertia and the momentum production of the system.
This equation is the same as that in Pudasaini and Fischer (2020) except
that the last term on the right hand side is now $2 u^b E$
instead of $u^b E$. In (\ref{Eqn_2_CON_B}) one $u^b E$ emerges from the momentum production
derived from the effectively reduced friction
 (as in Pudasaini and Fischer, 2020), while the other $u^b E$ originates from
the correct understanding of the inertia
of the entrained mass that has not yet been considered in any existing models including the one by Pudasaini and Fischer (2020). 
 However, mechanically and dynamically, this makes a huge difference, and thus, is a great advancement in simulating landslide with erosion.
Importantly, our present analysis makes a complete
description of the full dynamical model
equations for erosive landslide in conservative form by considering all
the aspects associated with the erosion-induced
reduced friction (the momentum production) and the correct handling of the inertia of the
system, which now reads:
\begin{equation}
\frac{\partial h}{\partial t} +  \frac{\partial }{\partial x}\left ( hu\right ) = E,
\label{Eqn_1_CON_C}
\end{equation}
\begin{eqnarray}
&&\frac{\partial}{\partial t}\left( hu \right)+ \frac{\partial}{\partial x} \left [ hu^2 + \left ( 1- \gamma^m\right)\alpha^m g^z K\frac{h^2}{2}\right]\nonumber\\ 
&&= h \left[g^x - \left ( 1-\gamma^m \right ) \alpha^m g^z\mu^m
-\left\{ 1-\left( 1-\gamma^m\right)\alpha^m\right\}g^z
\frac{\partial h}{\partial x} 
- C_{DV} u^2\right] + 2\lambda^b E u,
\label{Eqn_2_CON_C}
\end{eqnarray}
where, as before, the substitution $u^b = \lambda^b u$ has been made, and the dynamics of $\lambda^b \in (0, 1)$ has been
described in the previous sections.
All the analyses (with respect to $u^b$, or $\lambda^b$) presented above
about the gained or lost momentum (energy),
and the enhanced or reduced mobility of the erosive landslide are also
valid for the model equations (\ref{Eqn_1_CON_C})-(\ref{Eqn_2_CON_C})
written in conservative form and describe the evolution of the flow depth and the flow velocity. 

\subsection{Extension to multi-phase mass flow simulation}

It is now so convenient that the models (\ref{Eqn_1_CON_C})-(\ref{Eqn_2_CON_C}) can be directly applied to the mechanical erosion model for two-phase mass flows developed by Pudasaini and Fischer (2020) just by replacing $\lambda^b$ by $2\lambda^b$ in the momentum balance equations in Pudasaini and Fischer (2020). Importantly, this also lays a foundation to further apply these methods to multi-phase mass flow model (Pudasaini and Mergili, 2019) to include multi-phase erosion phenomena.  

\section{Discussions: Novelty, Essence and Implications}

Erosion-induced increased or decreased mobility has been reported with the relevant data in the laboratory and/or from the field events (Cascini et al., 2014; Cuomo et al., 2014, 2016; Pudasaini and Fischer, 2020). However, no clear-cut mechanical derivation and explanations have been presented in the literature so far for when and how the erosion related mass flow gains energy leading to enhanced-mobility, or loses energy that consequently forces the landslide to reduce its velocity and thus its run-out distance, resulting in reduced mobility. We have presented the first-ever, analytically constructed simple and clear mechanical condition for erosion-induced energy budget that delineates energy gain or loss, and the associated enhanced or reduced mass flow mobility.    
We analytically derived two important dimensionless numbers, the mobility scaling and the erosion number, as a function of the ratio between the erosion-induced force and the net driving force.  
These numbers provide the practitioners with the simple and direct measure of mobility by comparing the landslide dynamics without erosion from that with erosion. Furthermore, the mobility scaling offers an exact quantification of erosion in mobility.
\\[3mm]
By rigorous derivation, Pudasaini and Fischer (2020) showed that appropriate incorporation of the mass and momentum productions or losses in conservative model formulation is essential for the physically correct and mathematically consistent description of erosion-entrainment-deposition processes. They proved that effectively reduced friction in erosion is equivalent to the momentum production. With this, Pudasaini and Fischer (2020) demonstrated that erosion can induce a higher mass flow mobility. 
Although Pudasaini and Fischer (2020) laid the foundation for the mechanical erosion rate model, their model is incomplete as they did not deal with the erosion-induced inertia.  
In the conservative formulation, the change in inertia is implicit in the inertial part of the momentum balance equation. However, the explicit and full  quantification of the available energy (or, momentum), and thus the state of mobility of the erosive landslide, requires the combined analysis of the erosion-induced changed inertia of the system and the produced momentum. This is vital for applications in real world problems as it enables us to determine the actual change in the momentum and consequently the energy, and the mobility of the landslide with erosion. Here, we have achieved this by utilizing the non-conservative formulation which made it possible to explicitly express the changed inertia of the system that combines the erosion-related rate of change of mass (resulting in the mass production), $E$, and the relative velocity of the eroded particles with respect to the landslide velocity, namely the entrainment velocity, $u^{ev} = u - u^b$. This combination uniquely yielded the actual change in the inertia of the system, $\left ( u - u^b\right )E$. As shown above, this, together with the produced momentum, $u^bE$, provides the first-ever explicit and complete mechanical quantification of the (overall) state of induced momentum or energy, $\left(2\lambda^b -1 \right)E^P$, associated with the erosive landslide, and hence, the precise description of its mobility. This, in fact, fully addresses the long-standing scientific and engineering question of why and when some erosive landslides can have higher mobility, while others have their mobility reduced even in the erosive situation.   
\\[3mm]
Most of the existing and influential erosion models (Perla et al., 1980; Fraccarolla and Capart, 2002; McDougall and Hungr, 2005; Christen et al., 2010; Iverson, 2012; Frank et al., 2017) do not include the momentum production in the momentum balance equation. These aspects have also been partially discussed with a mechanical erosion model for two-phase mass flow by Pudasaini and Fischer (2020). Moreover, none of the existing models deal with erosion-induced changed inertia of landslide. 
Essentially, Perla et al. (1980), McDougall and Hungr (2005), Iverson (2012) directly replaced $\displaystyle{ \frac{d}{dt}\left ( m u\right)}$ with $\displaystyle{ m\frac{du}{dt} + u\frac{dm}{dt}}$ in the inertial part of the momentum equation which is mechanically invalid for the erosive landslide. Therefore, these types of models result in the unphysical loss of energy associated with the erosive landslide, and hence, cannot properly explain the state of energy and mobility.
\\[3mm]
So, from the physical point of view, most of the existing and dominant erosion models for mass movements are erroneous because of two reasons: First, incorrect formulation of the inertial part of the momentum equation. And, second, neglection of the momentum production, associated with the erosion velocity, in the momentum balance equation. Together with Pudasaini and Fischer (2020), we solved this fundamental problem in landslide motion with erosion paving now the way for the correct modelling and prediction of landslide dynamics, that erosion essentially changes the state of energy and mobility, and thus the run-out, dynamic impact and inundation.
So, the present contribution fundamentally enhances our understanding of mobility of erosive mass movements. 
\\[3mm]
We have also derived a full set of dynamical equations in conservative form in which the momentum balance correctly includes the erosion-induced change in inertia and the net energy gain or the net momentum production, $2\lambda^b E u$. This offers a legitimate simulation of landslide motion with erosion.
The new effectively single-phase mass flow model with erosion (\ref{Eqn_1_CON_C})-(\ref{Eqn_2_CON_C}) can be directly extended to crucially enhance the existing two-phase erosion model (Pudasaini and Fischer, 2020), or to the multi-phase mass flow model (Pudasaini and Mergili, 2019). This allows us to simulate much more realistic erosion related mass flow mobility in real events. The application of the novel model to experimental and complex natural events of landslides, debris and avalanche motions would require substantial additional work, and corresponding parameter estimates, either derived from field measurements or back calculations, involving observation data, which, therefore, has to be deferred to some future contributions.

\section{Summary}

Mobility of an erosive landslide is attributed to its excessive volume and material properties, and is marked by
rapid motion, exceptional travel distance and the inundation area. Proper knowledge of mobility is required in accurately determining
the dynamics and enormous impact
energy. However, most of the existing influential erosive landslide models do not include
momentum production and none of them deal correctly with
erosion-induced changed inertia. The correct consideration of the inertial
frame of reference is crucial for the precise derivation of the dynamical
landslide model with erosion. A novel understanding is that the increased  (changed) inertia is not related to the landslide velocity, but it is associated with the distinctly different entrainment velocity emerging
from the inertial frame of reference of the landslide. 
The classical representation of inertia appeared to be wrong for
erosional situations. We eliminated the existing erroneous perception on erosive
landslides and make a breakthrough in correctly determining the energy and
thus the mobility of the erosive landslide, that erosion fundamentally changes
the state of energy and mobility, and consequently the dynamic impact and
inundation. The actual change in inertia together with
the produced momentum provides the first-ever explicit and full mechanical
quantification of the state of energy, and thus, the precise
description of mobility of the erosive landslide. This addresses the
long-standing scientific question of why and when some
erosive landslides can have higher mobility, while the others have reduced
mobility.
\\[3mm]
We revealed, that the erosion velocity plays
an outstanding role in appropriately determining the energy budget and
mobility of an erosive landslide. The mobility is clearly defined and fully
controlled by the erosion velocity.
If the erosion velocity is greater than
one half of the flow velocity, the mobility is enhanced. Erosion velocity closer to the flow
velocity results in the most mobile flow. If the erosion velocity is less
than one half of the flow velocity then, the mobility is reduced. The landslide consumes the highest amount of
energy if the erosion velocity is much smaller than the flow velocity.
 The erosion will not change the energy status, and
hence the mobility of the landslide, if the erosion velocity is one half of
the flow velocity. Based on the newly constructed energy generator, we extracted an important
conclusion: whether the erosion related mass flow mobility will be
enhanced, reduced or remains unaltered depends exclusively on whether the
energy generator is positive, negative or zero. This becomes the game-changer and explicitly tells us
the state of mobility, and ultimately
regulates the destructive power of the landslide. With the
knowledge of the energy generator, the involved energy in landslide
erosion can now be quantified from the overall mobility parameter. Such an
explicit and fully mechanical description of the state of mobility of an erosive landslide is seminal. We introduced three principally novel mechanical concepts: the erosion-velocity,
entrainment-velocity and the energy-velocity, and demonstrated that the erosion and entrainment are
essentially different processes. With this, we drew a central
inference: the landslide gains energy, and thus enhances its mobility, if
the erosion velocity is greater than the entrainment velocity. We call
this phenomenon the landslide-propulsion, emerging from the net momentum
production, that provides the erosion-thrust to the landslide.
 The energy velocity associated with the net momentum production
 clearly delineates the three excess energy regimes: positive, negative, or zero, resulting in the corresponding enhanced, reduced or unaltered mobility of the erosional landslide.
\\[3mm]
Based on the newly derived basic erosional landslide equation, we
constructed a simple landslide mobility equation.
The great advantage of the new mobility equation is
that the erosion enhanced mobility can now be directly quantified.
This is the first-ever physics-based model to do so. 
We explicitly obtained the landslide velocity by analytically solving the
 mobility equation. 
This form of the velocity is very 
useful in quickly solving relevant engineering and applied problems and has enormous application potential. 
Analytically
obtained values well represent the velocity of natural landslides and debris
avalanches with erosion and demonstrate that erosion can have the major control on the landslide dynamics. As the
dynamic pressure is proportional to the square of the velocity, the
dynamic pressure with erosion can be much higher than the same without
erosion. Similarly, the large contrast in velocity with and without
erosion results in completely different run-out scenarios, much more extensive for erosive landslides.
Technically, this
provides a very important information for landslide practitioners in accurately determining the landslide velocity
with erosion.  
 We constructed two innovative and useful dimensionless numbers, the mobility scaling and erosion number, providing a direct measure of landslide mobility with erosion. This offers the unique possibility to precisely quantify the significance of erosion in mobility.
Importantly, we have also derived the full set of dynamical equations
 for landslide in conservative form
in which the momentum balance correctly includes the erosion
induced change in inertia and the momentum production.
 This is a great advancement in fully simulating landslide with
erosion. This clearly suggests the importance of the correct inclusion of
 erosion in modelling the landslide dynamics and run-out.

\section*{Acknowledgments} Shiva P. Pudasaini acknowledges the financial support provided by the Technical University of Munich with the Visiting Professorship Program, and the international research project: AlpSenseRely $-$ Alpine remote sensing of climate‐induced natural hazards - from the Bayerisches Staatsministerium f\"ur Umwelt und Verbraucherschutz, Munich, Bayern.


\begin{thebibliography}{99}

\bibitem{} Armanini, A., Fraccarollo, L., Rosatti, G., 2009. Two-dimensional
simulation of debris flows in erodible channels. Comput. Geosci. 35,
993-1006.

\bibitem{} Berger, C., McArdell, B.W., Schlunegger, F., 2011. Direct measurement of
channel erosion by debris flows, Illgraben, Switzerland. J. Geophys. Res.
Earth Surf. 116, F01002.

\bibitem{} Berzi, D., Fraccarollo, L., 2015. Turbulence locality and granularlike
fluid shear viscosity in collisional suspensions. Physical Review Letters
115, 194501.


\bibitem{} Brufau, P., Garcia-Navarro, P., Ghilardi, P., Natale, L., Savi, F., 2000. 1-d mathematical modelling of debris flow. J. Hydraul. Res. 38,
435-446.

\bibitem{} Cascini, L., Cuomo, S., Pastor, M., Rendina, I., 2016. SPH-FDM propagation
and pore water pressure modelling for debris flows in flume tests.
Engineering Geology 213, 74-83.

\bibitem{} Cascini, L., Cuomo, S., Pastor, M., Sorbino, G., Piciullo, L., 2014. SPH
run-out modelling of channelized landslides of the flow type.
Geomorphology 214, 502-513.

\bibitem{} Chen, H., Crosta, G.B., Lee, C.F., 2006. Erosional effects on runout of
fast landslides, debris flows and avalanches: A numerical investigation.
Geotechnique 56, 305-322.

\bibitem{} Christen, M., Kowalski, J., Bartelt, P., 2010. Ramms: numerical simulation
of dense snow avalanches in three-dimensional terrain. Cold Regions Science and Technology 63, 1-14.

\bibitem{} Crosta, G.B., Imposimato, S., Roddeman, D., 2009. Numerical modelling of
entrainment/deposition in rock and debris-avalanches. Eng. Geol. 109,
135-145.

\bibitem{} Cuomo, S., Pastor, M., Capobianco, V., Cascini, L., 2016. Modelling the
space time evolution of bed entrainment for flow-like landslides.
Engineering Geology 212, 10-20.

\bibitem{} Cuomo, S., Pastor, M., Cascini, L., Castorino, G.C., 2014. Interplay of
rheology and entrainment in debris avalanches: a numerical study. Canadian
Geotechnical Journal 51 (11), 1318-1330.

\bibitem{} de Haas, T., van Woerkom, T., 2016. Bed scour by debris flows:
experimental investigation of effects of debris flow composition. Earth
Surf. Process. Landforms 41, 1951-1966.


\bibitem{} de Haas, T., Nijland, W., de Jong, S.M., McArdell, B.W., 2020. How memory
effects, check dams, and channel geometry control erosion and deposition
by debris flows. Scientific Reports. 10, 14024.
https://doi.org/10.1038/s41598-020-71016-8.

\bibitem{} Dietrich, A., Krautblatter, M., 2019. Deciphering controls for debris-flow
erosion derived from a liDAR-recorded extreme event and a calibrated
numerical model (Rossbichelbach, Germany). Earth Surf. Process. Landform
44, 1346-1361.


\bibitem{} Dowling, C.A., Santi, P.M., 2014. Debris flows and their toll on human
life: a global analysis of debris-flow fatalities from 1950 to 2011. Nat.
Hazards 71(1), 203-227.

\bibitem{} Egashira, S., Honda, N., Itoh, T., 2001. Experimental study on the
entrainment of bed material into debris flow. Phys. Chem. Earth (C) 26,
645-650.

\bibitem{} Erismann, T.H., Abele, G., 2001. Dynamics of Rockslides and Rockfalls. Springer, New York.

\bibitem{} Evans, S.G., Bishop, N.F., Fidel Smoll, L., Valderrama Murillo, P.,
Delaney, K.B., Oliver-Smith, A., 2009. A re-examination of the mechanism
and human impact of catastrophic mass flows originating on Nevado
Huascaran, Cordillera Blanca, Peru in 1962 and 1970. Eng. Geol. 108,
96-118.


\bibitem{} Fagents, S.M., Baloga, S.A., 2006. Toward a model for the bulking and
debulking of lahars. J. Geophys. Res. 111, B10201. doi:10.1029/2005JB003986.

\bibitem{} Faug, T., 2015. Depth-averaged analytic solutions for free-surface
granular flows impacting rigid walls down inclines. Phys. Rev. E
92. http://dx.doi.org/10.1103/PhysRevE.92.062310.

\bibitem{} Faug, T., Chanut, B., Beguin, R., Naaim, M., Thibert, E., Baraudi, D.,
2010. A simple analytical model for pressure on obstacles induced by snow
avalanches. Ann. Glaciol. 51 (54), 1-8.

\bibitem{} Fraccarollo, L., Capart, H., 2002. Riemann wave description of erosional
dam-break flows. J. Fluid Mech. 461, 183-228.

\bibitem{} Frank, F., McArdell, B.W., Huggel, C., Vieli, A., 2015. The importance of
entrainment and bulking on debris flow runout modeling: examples from the
Swiss Alps. Nat. Hazards Earth Syst. Sci. 15, 2569-2583.

\bibitem{} Gubler, H., 1989. Comparison of three models of avalanche dynamics. Annals
of Glaciology 13, 82-89.

\bibitem{} Highland, L.M., Bobrowsky, P., 2008. The landslide handbook - A guide to understanding landslides: Reston, Virginia, U.S. Geological Survey Circular 1325, 129 p.

\bibitem{} Huggel, C., Zgraggen-Oswald, S., Haeberli, W., K\"a\"ab, A., Polkvoj, A.,
Galushkin, I., Evans, S.G., 2005. The 2002 rock/ice avalanche at
Kolka/Karmadon, Russian Caucasus: assessment of extraordinary avalanche
formation and mobility, and application of QuickBird satellite imagery.
Nat. Hazards Earth Syst. Sci. 5, 173-187.

\bibitem{} Hungr, O., McDougall, S., Bovis, M., 2005. Entrainment of material by
debris flows. In Debris-flow hazards and related phenomena, Eds. Jakob, M.
\& Hungr, O., Springer, Berlin.


\bibitem{} Iverson, R. M., Ouyang, C., 2015. Entrainment of bed material by
earth-surface mass flows: review and reformulation of depth-integrated
theory. Rev. Geophys. 53(1), 27-58.

\bibitem{} Iverson, R.M., 2012. Elementary theory of bed-sediment entrainment by
debris flows and avalanches. J. Geophys. Res. 117, F03006.

\bibitem{} Iverson, R., Reid, M., Logan, M. et al., 2011. Positive feedback and
momentum growth during debris-flow entrainment of wet bed sediment. Nature
Geosci. 4, 116-121. 

\bibitem{} Jenkins, J.T., Berzi, D., 2016. Erosion and deposition in depth-averaged
models of dense, dry, inclined, granular flows. Physical Review E 94,
052904.


\bibitem{} Kafle, J., Kattel, P., Mergili, M., Fischer, J.-T., Pudasaini, S.P., 2019.
Dynamic response of submarine obstacles to two-phase landslide and tsunami
impact on reservoirs. Acta Mech 230, 3143-3169.

\bibitem{} Kattel, P. , Kafle, J. , Fischer, J.-T. , Mergili, M. , Tuladhar, B.M.,
Pudasaini, S.P., 2018. Interaction of two-phase debris flow with
obstacles. Eng. Geol. 242, 197-217.

\bibitem{} Lanzoni, S., Gregoretti, C., Stancanelli, L.M., 2017. Coarse-grained
debris flow dynamics on erodible beds. J. Geophys. Res. Earth Surf.
122(3), 592-614.

\bibitem{} Le, L., Pitman, E.B., 2009. A model for granular flows over an erodible
surface. SIAM J. Appl. Math. 70, 1407-1427.

\bibitem{} Li, P., Hu, K., Wang, X., 2017. Debris flow entrainment rates in
non-uniform channels with convex and concave slopes. J. Hydraul. Res. 56,
1-12.

\bibitem{} Li, P., Shen, W., Hou, X., Li, T., 2019. Numerical simulation of the
propagation process of a rapid flow-like landslide considering bed
entrainment: A case study. Engineering Geology 263, 105287.
doi:10.1016/j.enggeo.2019.105287.

\bibitem{} Liu, W., Yang, Z., He, S., 2021. Modeling the landslide-generated debris flow from formation to propagation and run-out by considering the effect of vegetation. Landslides 18, 43–58. 

\bibitem{} Liu, W., He, S., 2020. Comprehensive modelling of runoff-generated debris flow from formation to propagation in a catchment. Landslides. doi:10.1007/s10346-020-01383-w.

\bibitem{} Liu, W., Wang, D., Zhou, J., He, S., 2019. Simulating the Xinmo landslide runout considering entrainment effect. Environ. Earth Sci. 78, 585. doi:10.1007/s12665-019-8596-2.

\bibitem{} Lu, P.Y., Yang, X.G., Xu, F.G., Hou, T.X., Zhou, J.W., 2016. An
analysis of the entrainment effect of dry debris avalanches on loose bed
materials. SpringerPlus 5(1), 1621.

\bibitem{} McClung, D. M., 1983. Derivation of Voellmy’s Maximum Speed and Run-Out
Estimates from a Centre-of-Mass Model. Journal of Glaciology 29(102), 350-352. 

\bibitem{} McDougall, S., Hungr, O., 2005. Dynamic modelling of entrainment in rapid
landslides. Can. Geotech. J. 42, 1437-1448.

\bibitem{} McCoy, S.W., Kean, J.W., Coe, J.A., Tucker, G.E., Staley, D.M.,
Wasklewicz, T.A., 2012. Sediment entrainment by debris flows: In situ
measurements from the headwaters of a steep catchment. J. Geophys. Res.
117, F03016. doi:10.1029/2011JF002278.

\bibitem{} Medina, V., H\"urlimann, M., Bateman, A., 2008. Application of FLATModel, a
2D finite volume code, to debris flows in the northeastern part of the
Iberian Peninsula. Landslides 5, 127-142.

\bibitem{} Mergili, M., Jaboyedoff, M., Pullarello, J., Pudasaini, S.P., 2020b. Back
calculation of the 2017 Piz Cengalo - Bondo landslide cascade with
r.avaflow: what we can do and what we can learn. Nat. Hazards Earth Syst.
Sci. 20, 505-520.

\bibitem{} Mergili, M., Pudasaini, S.P., Emmer, A., Fischer, J.-T., Cochachin, A.,
Frey, H., 2020a. Reconstruction of the 1941 GLOF process chain at lake
Palcacocha (Cordillera Blanca, Peru). Hydrol. Earth Syst. Sci. 24,
93-114.

\bibitem{} Mergili, M., Emmer, A., Juricova, A., Cochachin, A., Fischer, J.-T.,
Huggel, C., Pudasaini, S.P., 2018. How well can we simulate complex
hydro-geomorphic process chains? The 2012 multi-lake outburst flood in the
Santa Cruz Valley (Cordillera Blanca, Peru). Earth Surf. Proc. Land. 43,
1373-1389.

\bibitem{} Naaim, M., Faug, T., Naaim-Bouvet, F., 2003. Dry granular flow modeling
including erosion and deposition. Surv. Geophys. 24, 569-585.

\bibitem{} Nikooei, M., Manzari, M.T., 2020. Studying effect of entrainment on
dynamics of debris flows using numerical simulation. Computers and
Geosciences 134, 104337. doi:10.1016/j.cageo.2019.104337.

\bibitem{} Perla, R., Cheng, T.T., McClung, D.M., 1980. A two-parameter model for
snow-avalanche motion. J. Glaciology 26(94), 197-207.

\bibitem{} Pilvar, M., Pouraghniaei, M.J., Shakibaeinia, A., 2019. Two-dimensional subaerial, submerged, and transitional granular slides. Physics of Fluids 31, 113303. https://doi.org/10.1063/1.5121881.

\bibitem{} Pudasaini, S.P., Krautblatter, M., 2021. The Landslide Velocity. arXiv:2103.10939.

\bibitem{} Pudasaini, S.P., Fischer, J.-T., 2020. A mechanical erosion model for
two-phase mass flows. International Journal of Multiphase Flow 132, 103416. https://doi.org/10.1016/j.ijmultiphaseflow.2020.103416.

\bibitem{} Pudasaini, S.P., Mergili, M., 2019. A multi-phase mass flow model. Journal of Geophysical Research: Earth Surface 124, 2920-2942. 

\bibitem{} Pudasaini, S.P., Ghosh Hajra, S., Kandel, S., Khattri, K.B., 2018. Analytical
solutions to a nonlinear diffusion-advection equation. Zeitschrift f\"ur
angewandte Mathematik und Physik 69(6), 150.
https://doi.org/10.1007/s00033-018-1042-6.

\bibitem{} Pudasaini, S.P., 2016. A novel description of fluid flow in porous and debris materials. Eng. Geol. 202, 62-73.

\bibitem{} Pudasaini, S.P., Krautblatter, M., 2014. A two-phase mechanical model for
rock-ice avalanches. J. Geophys. Res. Earth Surf. 119. doi:10.1002/2014JF0 03183.

\bibitem{} Pudasaini, S.P., 2012. A general two-phase debris flow model. J.
Geophysics. Res. 117, F03010. doi:10.1029/2011JF002186.

\bibitem{} Pudasaini, S.P., 2011. Some exact solutions for debris and avalanche
flows. Phys. Fluids 23, 043301. doi:10.1063/1.3570532.

\bibitem{} Pudasaini, S.P., Hutter, K., 2007. Avalanche Dynamics: Dynamics of Rapid
Flows of Dense Granular Avalanches. Springer, Berlin, New York.

\bibitem{} Qiao, C., Ou, G., Pan, H., 2019. Numerical modelling of the long runout
character of 2015 Shenzhen landslide with a general two-phase mass flow
model. Bull. Eng. Geol. Environ. 78, 3281-3294.

\bibitem{} Reid, M.E., Iverson, R.M., Logan, M., Lahusen, R.G., Godt, J.W., Griswold,
J.P., 2011. Entrainment of bed sediment by debris flows: results from
large-scale experiments. Italian J. Eng. Geol. Env.
doi:10.4408/IJEGE.2011-03.B-042.

\bibitem{} Rickenmann, D., Weber, D., Stepanov, B., 2003. Erosion by debris flows in
field and laboratory experiments. In: Debris-Flow Hazards Mitigation:
Mechanics, Prediction, and Assessment, Ed. Rickenmann, D., Chen, C.,
883-894. Millpress, Rotterdam, Netherlands.

\bibitem{} Rickenmann, D., 1999. Empirical relationships for debris flows. Nat.
Hazards 19(1), 47-77.

\bibitem{} Salm, B., 1966, Contribution to avalanche dynamics. International
Symposium on Scientific Aspects of Snow and Ice Avalanches, 1965, Davos;
pp. 199-214: IAHS Publ. No. 69.

\bibitem{} Santi, P.M., Higgins, J.D., Cannon, S.H., Gartner, J.E., 2008. Sources of
debris flow material in burned areas. Geomorphology 96, 310-321.


\bibitem{} Sch\"urch, P., Densmore, A.L., Rosser, N.J., McArdell, B.W., 2011.
Dynamic controls on erosion and deposition on debris-flow fans. Geology
39(9), 827-830.

\bibitem{} Shen, W., Li, T., Li, P., Berti, M., Shen, Y., Guo, J., 2019. Two-layer
numerical model for simulating the frontal plowing phenomenon of flow-like
landslides. Engineering Geology 259, 105168. doi:10.1016/j.enggeo.2019.105168 .


\bibitem{} Sokolsky, V.N., 1968. K. E. Tsiolkovsky Selected Works, MIR Publishers (Moscow), English Translation.

\bibitem{} Somos-Valenzuela, M.A., Chisolm, R.E., Rivas, D.S., Portocarrero, C.,
McKinney, D.C., 2016. Modeling a glacial lake outburst flood process
chain: the case of Lake Palcacocha and Huaraz, Peru. Hydrol. Earth Syst.
Sci. 20, 2519–2543.

\bibitem{} Sovilla, B., Burlando, P., Bartelt, P., 2006. Field experiments and
numerical modeling of mass entrainment in snow avalanches. J. Geophys.
Res. 111, F03007.

\bibitem{} Tai, Y.-C., Noelle, S., Gray, J.M.N.T., Hutter, K., 2002. Shock-capturing
and front-tracking methods for granular avalanches. J. Comput. Phys. 175(1), 269-301.

\bibitem{} Tai, Y.-C., Kuo, C.Y., 2008. A new model of granular flows over general
topography with erosion and deposition. Acta Mech. 199, 71-96.

\bibitem{} Takahashi, T., Kuang, S.F., 1986. Formation of debris flow on varied slope
bed. Disas. Prev. Res. Inst. Annuals B 29, 343-359.

\bibitem{} Theule, J. I., Liebault, F., Laigle, D., Loye, A., Jaboyedoff, M., 2015.
Channel scour and fill by debris flows and bedload transport.
Geomorphology 243, 92-105.

\bibitem{} Voellmy. A., 1955. \"Uber die Zerst\"orungskraft von Lawinen.
Schweizerische Bauzeitung. Jahrg. 73. Ht. 12., 159-162; Ht. 15, 212-217;
Ht. 17, 246-249; Ht. 19, 280-285. On the destructive force of avalanches,
Translation No. 2. Alta. Avalanche Study Center, USDA, Forest Service,
1964.

\bibitem{}  Walter, F., Amann, S., Kos, A., Kenner. R., Phillips, M., de Preux, A., Huss, M., Tognacca, C., Clinton, J., Diehl, T., Bonanomi, Y., 2020. Direct observations of a three million cubic meter rock‐slope collapse with almost immediate initiation of ensuing debris flows. Geomorphology 351, 106933. https://doi.org/10.1016/j.geomorph.2019.106933.


\end{thebibliography}
\end{document}